\documentclass[fleqn,usenatbib]{mnras}

\usepackage[T1]{fontenc}
\usepackage{ae,aecompl}

\usepackage{graphicx}	% Including figure files
\usepackage{amsmath}	% Advanced maths commands
\usepackage{multicol}   % Multi-column entries in tables
\usepackage{bm}		    % Bold maths symbols, including upright Greek
\usepackage{pdflscape}	% Landscape pages
\usepackage{epsfig,color,pifont}
\usepackage{times}
\usepackage{subfig}
\usepackage{url}

\usepackage{graphicx}	% Including figure files
\usepackage{amsmath}	% Advanced maths commands
\usepackage{amssymb}	% Extra maths symbols

\usepackage{newtxtext,newtxmath}

\def\Msun{\mbox{~M$_\odot$}}

\def\Msunpc2{\mbox{~M$_\odot$~pc$^{-2}$}}
\def\Msunyr{\mbox{~M$_\odot$~yr$^{-1}$}}
\def\kms{\mbox{~km~s$^{-1}$}}

\def\kpc{\mbox{~kpc}}
\def\Mpc{\mbox{~Mpc}}

\def\Gyr{\mbox{~Gyr}}
\def\erg{\mbox{~erg}}

\def\SigmaSFR{\Sigma_{\rm SFR}}
\def\Sigmagas{\Sigma_{\rm gas}}

\def\fbound{f_{\rm bound}}

\def\Mmin{M_{\rm min}}
\def\Mmax{M_{\rm max}}

\def\sigmac{\sigma_{\rm c}}

\def\rhogas{\rho_{\rm gas}}
\def\phiP{\phi_P}
\def\sigmadyn{\sigma_{\rm dyn}}

\def\fH2{f_{\rm H2}}
\def\sigmaHI{\sigma_{\rm HI}}
\def\sigmaH2{\sigma_{\rm H2}}
\def\csHI{c_{\rm s,HI}}
\def\csH2{c_{\rm s,H2}}

\def\vcirc{V_{\rm circ}}
\def\vmax{V_{\rm max}}

\def\vout{V_{out}}

\def\Vmax{V_{\rm max}}

\def\vout{V_{\rm out}}

\def\vbar{V_{\rm bar}}
\def\vDM{V_{\rm DM}}
\def\fbar{f_{bar}}

\def\R200{R_{200}}
\def\M200{M_{200}}

\def\MDM{M_{\rm DM}}
\def\Mbar{M_{\rm bar}}
\def\Mdyn{M_{\rm dyn}}
\def\fbar{f_{\rm bar}}
\def\Mgas{M_{\rm gas}}
\def\Mstar{M_*}
\def\MHI{M_{\rm HI}}
\def\Mstarmean{\langle \Mstar \rangle}
\def\Mstaryoung{\Mstar^{\rm{young}}}
\def\Mhalo{M_{\rm halo}}

\def\Omegabar{\Omega_{\rm bar}}
\def\Omegamatter{\Omega_{\rm m}}
\def\rs{r_{\rm s}}
\def\rhos{\rho_{\rm s}}
\def\rhocrit{\rho_{\rm crit}}
\def\acoll{a_{\rm coll}}
\def\zcoll{z_{\rm coll}}
\def\tcoll{t_{\rm coll}}
\def\sigmac{\sigma_{\log c}}
\def\SFR{{\rm SFR}}
\def\Msdot{\dot{M}_*}
\def\Ncl{N_{\rm cl}^{\zcoll}}
\def\NGC{N_{\rm GC}^{\zcoll}}
\def\TN{T_N^{\zcoll}}
\def\diff{{\rm d}}
\def\fcl{n_{\rm cl, SN}}
\def\etaw{\eta_{\rm w}}
\def\etagal{\eta_{\rm gal}}
\def\fDM{f_{\rm DM}}
\def\Mcl{M_{\rm cl}^{\zcoll}}
\def\Mclmean{\langle\Mcl\rangle}
\def\MSN{M_{\rm SN}}
\def\rcore{r_{\rm core}}
\def\tSF{t_{\rm SF}}
\def\tdyn{t_{\rm dyn}}
\def\etaDM{\eta_{\rm DM}}
\def\NSN{N_{\rm SN}}
\def\ESN{E_{\rm SN}}
\def\McNFW{M_{\rm cNFW}}

\def\re{r_{\rm e}}
\def\reyoung{r_{\rm e}^{\rm young}}
\def\reold{r_{\rm e}^{\rm old}}
\def\regas{r_{\rm e,gas}}
\def\fgas{f_{\rm gas}}
\def\fc{f_c}
\def\fstar{f_*}
\def\rout{r_{\rm out}}
\def\Mgc{M_{\rm GC}}

\def\apj{\rm Astrophys. J.}
\def\aap{\rm Astron. \&Astrophys.}
\def\mnras{\rm Mon.\,Not.\,RAS.}
\def\LCDM{$\Lambda$CDM}
\def\Msunh{h^{-1}\mbox{M}_\odot}

\def\mathnew{\mathsurround=0pt}
\def\simov#1#2{\lower .5pt\vbox{\baselineskip0pt
    \lineskip-.5pt\ialign{$\mathnew#1\hfil##\hfil$\crcr#2\crcr\sim\crcr}}}

\def\apj{\rm ApJ}
\def\apjl{\rm ApJL}
\def\apjs{\rm ApJS}
\def\aj{\rm AJ}
\def\mnras{\rm MNRAS}
\def\nat{\rm Nature}

\def\aap{\rm A\&A}

\def\pasp{\rm PASP}

\newcommand{\HI}{\hbox{{\sc H}\hspace{0.7pt}{\sc i}}}

\defcitealias{Reina-Campos17}{RK17}
\defcitealias{Kruijssen12d}{K12}
\defcitealias{Trujillo-Gomez19}{TRK19}

\title[UDG formation and feedback from GCs]{The emergence of dark matter-deficient ultra-diffuse galaxies driven by scatter in the stellar mass-halo mass relation and feedback from globular clusters}

\author[S. Trujillo-Gomez et al.]
{Sebastian Trujillo-Gomez$^{1}$\thanks{E-mail: strujill@gmail.com}, J.~M.~Diederik Kruijssen$^{1}$, 
and Marta Reina-Campos$^{1,2,3}$
\\
$^{1}$Astronomisches Rechen-Institut, Zentrum f{\"u}r Astronomie der Universit{\"a}t Heidelberg, Monchhofstra{\ss}e 12-14, D-69120 Heidelberg, Germany \\
$^{2}$ Department of Physics \& Astronomy, McMaster University, 1280 Main Street West, Hamilton, L8S 4M1, Canada\\
$^{3}$Canadian Institute for Theoretical Astrophysics (CITA), University of Toronto, 60 St George St, Toronto, M5S 3H8, Canada
}

\date{Accepted 2021 November 5. Received 2021 November 4; in original form 2021 March 12}

\pubyear{2021}

\begin{document}
\label{firstpage}
\pagerange{\pageref{firstpage}--\pageref{lastpage}}
\maketitle

% Abstract of the paper. The abstract should briefly describe the aims, methods, and main results of the paper. It should be a single paragraph not more than 250 words (200 words for Letters). No references should appear in the abstract.
\begin{abstract}
In addition to their low stellar densities, ultra-diffuse galaxies (UDGs) have a broad variety of dynamical mass-to-light ratios, ranging from dark matter (DM) dominated systems to objects nearly devoid of DM. To investigate the origin of this diversity, we develop a simple, semi-empirical model that predicts the structural evolution of galaxies, driven by feedback from massive star clusters, as a function of their departure from the mean SMHM relation. The model predicts that a galaxy located $\gtrsim 0.5$ dex above the mean relation at $\Mhalo=10^{10}\Msun$ will host a factor of $\sim 10-100$ larger globular cluster (GC) populations, and that feedback from these GCs drives a significant expansion of the stellar component and loss of DM compared to galaxies on the SMHM relation. This effect is stronger in haloes that collapse earlier and have enhanced star formation rates at $z\ga2$, which leads to increased gas pressures, stellar clustering, and mean cluster masses, and significantly enhances the energy loading of galactic winds and its impact on the DM and stellar orbits. The impact on galaxy size and DM content can be large enough to explain observed galaxies that contain nearly the universal baryon fraction, as well as NGC1052-DF2 and DF4 and other isolated UDGs that contain almost no DM. The trend of increasing galaxy size with GC specific frequency observed in galaxy clusters also emerges naturally in the model. Our predictions can be tested with large and deep surveys of the stellar and GC populations in dwarfs and UDGs. Because stellar clustering drives the efficiency of galactic winds, it may be a dominant factor in the structural evolution of galaxies and should be included as an essential ingredient in galaxy formation models. 
\end{abstract}

% Select between one and six entries from the list of approved keywords.
% Don't make up new ones.

%%%%%%%%%%%%%%%%%%%%%%%%%%%%%%%%%%%%%%%%%%%%%%%%%%

%%%%%%%%%%%%%%%%% BODY OF PAPER %%%%%%%%%%%%%%%%%%

\section{Introduction}
\label{sec:intro}

Over the last few years, interest in the study of so-called ultra-diffuse galaxies (UDGs) has rapidly increased. UDGs are typically defined to be galaxies with the stellar mass of a dwarf ($10^7 \la \Mstar/\Msun \la 3\times10^8$), and very large extent, $\re>1.5\kpc$, similar to the sizes of $L^*$ galaxies \citep{vanDokkum15}. Due to their very low surface brightness $\mu_e>24~{\rm mag~arcsec}^{-2}$, these galaxies were mostly absent in large surveys like SDSS\footnote{Sloan Digital Sky Survey, https://www.sdss.org}. When going to lower brightness limits, UDGs have been found in large numbers in galaxy clusters \citep{Caldwell06,vanDokkum15,Koda15, Mihos15,vandenBurg16}, but also in groups \citep{Merritt16,Trujillo17,RomanTrujillo17b}, and in the field \citep{Leisman17}, which suggest that they may represent a substantial fraction of the galaxy population \citep{Jones18}. 

Much of the interest in these objects is focused on establishing the mechanisms that lead to their formation, as well as how they fit, alongside normal galaxies, into the broader theory of galaxy formation in the context of the cold dark matter (\LCDM) cosmology. Many studies of their properties find that most UDGs populate dark matter (DM) haloes comparable to those hosting dwarf galaxies \citep[$\Mhalo \la 10^{11}\Msun$;][]{Beasley16,BeasleyTrujillo16,PengLim16,RomanTrujillo17a,Leisman17,Amorisco18a,Amorisco18b,Chilingarian19}. Among the current theoretical and numerical efforts to understand their formation, isolated UDGs have been found to  naturally arise in some cosmological hydrodynamical simulations as a result of feedback-driven winds that expand the dark matter and stellar orbits \citep{DiCintio17,Jackson20a}. Other models predict that UDGs form in the high-spin tail of the distribution of DM haloes obtained from cosmological simulations \citep{AmoriscoLoeb16}. However, not all of the observed UDGs can be explained by current models.

Recent studies have uncovered UDGs with surprising properties. \citet{vanDokkum18b} and \citet{vanDokkum19} found two UDGs in a nearby galaxy group, NGC1052-DF2 and DF4, hosting populations of $\sim 10$ unusually massive GCs ($\widetilde{M_{\rm GC}} \approx 9\times10^5\Msun$). Both galaxies have stellar masses $\sim 10^8\Msun$ and seem to contain very little DM, with stars providing nearly the entire dynamical mass in the central $\sim 8\kpc$ \citep{vanDokkum18a,vanDokkum19}. For an NFW DM halo density profile \citep{Navarro97}, this would imply a baryon fraction $\fbar \equiv \Mbar/\Mdyn \ga 50$ per cent, in contradiction with the $\la 3$ per cent expected in galaxies of the same mass in \LCDM ~\citep{Papastergis12}. Many theoretical studies have focused on understanding how DM-deficient galaxies could form in a Universe dominated by DM, pointing to the need for intense ram pressure and tidal stripping of satellites in highly eccentric orbits to preferentially remove the DM \citep{Ogiya18,Carleton18,Jiang19,Nusser20,Sales20,Maccio20,Jackson20b}. These results seem are in general very sensitive to the infall times and orbits of the galaxies.

UDGs found in isolation seem to present a greater challenge to current models. These are usually $\HI$-rich and star-forming, are commonly found in blind $\HI$ surveys \citep{Leisman17}, and have lower gas rotation rates than dwarfs of similar mass \citep{Jones18}. \citet{ManceraPina19,ManceraPina20} 
studied the $\HI$ rotation curves of six of these objects and found that they imply very high baryon fractions of $>47$ per cent (or conversely, very low DM content) within the extent of their gas discs, $\sim8-10\kpc$. These values are well above the budget defined by the cosmic mean, $\fbar\approx15$ per cent \citep{Planck15}. In the most extreme cases, the baryonic mass alone is enough to account for the rotation velocity of the gas. Since environmental processes cannot account for the large sizes and low DM content of these isolated systems, an internal mechanism is necessary. However, none of the mechanisms suggested so far for forming field UDGs, such as high spin DM haloes \citep{AmoriscoLoeb16} or feedback-driven expansion \citep{DiCintio17,Jiang19,Jackson20a}, are able to explain the extreme DM deficiency of these of objects.

The relation between the stellar mass of a galaxy and that of its host DM halo is at the heart of the problem of galaxy formation in the context of the current cosmological paradigm. The stellar mass-halo mass (SMHM) relation has been studied extensively in observations using many techniques \cite[e.g.][]{Leauthaud12,Behroozi19}. The halo abundance matching (HAM) technique combined with observed galaxy clustering \citep{Conroy06,Trujillo-Gomez11,Reddick13,Campbell18}provide useful constraints on the mean and scatter of the relation for massive galaxies \citep{tasitsiomi04,Kravtsov04,Conroy06,Moster10,Behroozi10,Rodriguez-Puebla12,Moster18}. However, at the scales of dwarf galaxies ($\Mhalo \la 10^{11}\Msun$), independent statistical constraints on the scatter in the SMHM relation, such as galaxy clustering or weak lensing, are not yet available. Dynamical modelling of the stellar and gaseous components of dwarf galaxies provide estimates of the DM halo mass, but these are often subject to significant systematics, including extrapolation from the galaxy extent out to the virial radius \citep{Read16b,Campbell17,Oman19}. Taken at face value, these estimates show broad agreement with the mean HAM expectation, but with considerably larger scatter in stellar mass at fixed halo mass \citep{Katz17,Read17,Schneider17,Forbes18b,Li20}.

In addition to observational evidence, larger scatter is also expected on theoretical grounds at the scales of dwarf galaxies due to their shallower potential wells, and the strong role of feedback and reionisation in the stochasticity of their star formation. These considerations raise the interesting possibility that dwarf galaxies could scatter around the mean SMHM relation by more than a decade in stellar mass at fixed halo mass, and that this scatter could have important consequences for the evolution of galaxies. In this work we develop a semi-empirical model to investigate the effect of this scatter on the structure, evolution, and star cluster populations of present day low-mass galaxies. The model relates the collapse epoch of a DM halo with its excursion from the mean SMHM relation, and predicts how this determines its star and cluster formation rate at early times, and how SN feedback from the enhanced stellar clustering modifies the structure of the galaxy and its DM halo at present.
  
This paper is organised as follows. Section~\ref{sec:model} describes the semi-empirical model. The predictions for galaxy and DM halo structure and GC populations at $z=0$ are presented in Section ~\ref{sec:global}, including the implications for the origin of UDGs, and how DM-deficient galaxies may be explained as a result of upward scatter in the SMHM relation. Section~\ref{sec:obs_halomass} shows the effect of structural evolution on the inferred halo masses of low-mass galaxies,  Section~\ref{sec:discussion} discusses the implications of the model, and Section~\ref{sec:conclusions} summarises our findings. The analysis below assumes the {\it Planck} cosmological parameters \citep{Planck15}: $\Omegamatter=0.308$, $\Omegabar=0.049$, $\sigma_8=0.816$, and $H_0=67.8\kms\Mpc^{-1}$.

\section{Modelling the impact of scatter in the SMHM relation on galaxy evolution and structure}
\label{sec:model}

In this section we describe a semi-empirical model that relates the departure of a galaxy from the SMHM relation to the evolution of its structure. In the model, both the integrated properties and the evolution of the galaxy are defined by its position in the 2-dimensional plane defined by $\Mstar$ and $\Mhalo$. The model is summarised as follows:
\begin{enumerate}
    \item For a given DM halo mass $\Mhalo$, the mean stellar mass at $z=0$ is determined by assuming an empirical mean SMHM relation.
    \item The scatter in $\Mstar$ at fixed $\Mhalo$ is assumed to correlate with the scatter in maximum circular velocity $\Vmax$ at fixed $\Mhalo$, such that more concentrated galaxies (i.e. with deeper potential wells and higher $\Vmax$) have larger $\Mstar$ at $z=0$.
    \item The DM halo concentration (now a function of both $\Mstar$ and $\Mhalo$) is used to obtain the halo collapse time through the direct relation found in cosmological simulations. This implies that galaxies with an excess of $\Mstar$ at a fixed $\Mhalo$ will inhabit denser DM haloes with earlier formation times relative to those on the mean SMHM relation. 
    \item Galaxies are assumed to have formed a fixed fraction of their present stellar mass during the collapse epoch of their DM halo. Gas accretion rates are by definition higher during halo collapse, leading to elevated gas pressures and star formation rates. Hence, this will correspond to the dominant epoch of GC formation in the galaxy.
    \item At the collapse time, sizes are assigned to the stellar component of galaxies based on an empirical relation between the effective radius $\re$ and $\Mhalo$ at that epoch. Using the size and stellar mass at the collapse time, the time-averaged SFR surface density within $\re$ is obtained.
    \item The gas surface density $\Sigmagas$ is calculated from the SFR surface density $\SigmaSFR$ at the time of collapse using the Kennicutt-Schmidt \citep{Kennicutt98} relation. The galactic rotation frequency $\Omega$ is obtained using the DM, stellar, and gas masses enclosed within $\re$. A model describing the dependence of the clustering of star formation on the galactic environment (specified by $\Sigmagas$, $\Omega$, and Toomre $Q$) is used to estimate the fraction of star formation occurring in bound clusters. A complementary model is applied to predict the environmentally-dependent initial cluster mass function (ICMF). The stellar cluster populations are fully determined by the bound fraction and the ICMF, including the number of GCs and the mean cluster mass.
    \item Using the results of detailed numerical simulations, the effect of supernovae (SN) clustering on the energy loading of galactic winds is estimated. The SN clustering is obtained from the SFR, the bound fraction of star formation, and the ICMF.
    \item To model the process of DM core formation due to feedback-driven outflows, a fixed fraction of the wind energy is assumed to couple to the DM halo potential energy at $z=0$. The model then predicts the amount of expansion of the collisionless components (DM and old stars) due to the increase in halo potential energy from galactic winds. This is done using a parametrized feedback-modified cored DM density profile calibrated using hydrodynamic simulations.  
    \item The masses and sizes of the DM and stellar components at the collapse time, and at $z=0$ derived above fully specify the structural evolution of the galaxy and define the dynamical mass of each component at all radii.
\end{enumerate}

The model outlined above thus connects the position of a galaxy in the $\Mstar - \Mhalo$ plane to its mean star formation rate density, cold gas content, massive star cluster populations, galactic winds, and feedback-driven structural evolution of its DM and stellar components. We have deliberately chosen to neglect the effect of individual mergers because these are difficult to model analytically and because we are primarily interested in the evolution of the galaxy and its GCs at $z\ga2$. At this epoch, smooth gas accretion from the cosmic web is the dominant mechanism for the production of massive clusters in the progenitors of $L^*$ galaxies \citep{Keller20}. Furthermore, an individual massive merger can be thought of as a temporary enhancement in the gas, stellar, and DM mass accretion rate that boosts the conversion of gas into stars, effectively scattering the remnant galaxy upwards from the mean SMHM relation. In a broad statistical sense, using mean halo mass accretion histories, our model then accounts for the effects of mergers. Furthermore, we do not account for GCs accreted from mergers, but these are expected to be subdominant in galaxies with $\Mhalo<10^{11}\Msun$ \citep{Choksi19}. In the following subsections we describe each component of the model in detail.

\subsection{The SMHM relation}

The starting point is the assumption of a mean SMHM relation at $z=0$. We employ the SMHM relation as derived by \citet{Behroozi13} using the halo abundance matching technique (HAM) and considering the stellar mass function, specific star formation rates, and the cosmic history of star formation as constraints. 

HAM generally consists of fitting the function $\Mstar(\Mhalo)$ by statistically matching the number density of observed galaxies with the DM halo mass function predicted by \LCDM. The low mass end of the SMHM relation, $\Mhalo \la 10^{11}\Msun$, is subject to significant systematics due to uncertainties in the stellar mass function, and the lack of constraints from galaxy clustering at low masses \citep[e.g. see figure 34 in][]{Behroozi19}. Furthermore, the relation extends down to $\Mhalo \simeq 10^{10}\Msun$, and extrapolation is required below this mass. Assuming a different mean SMHM relation would modify the quantitative predictions of our model, but not the overall behaviour of galaxies that scatter above (and below) compared to those on the mean relation.

\begin{figure*}
    \includegraphics[width=0.65\textwidth]{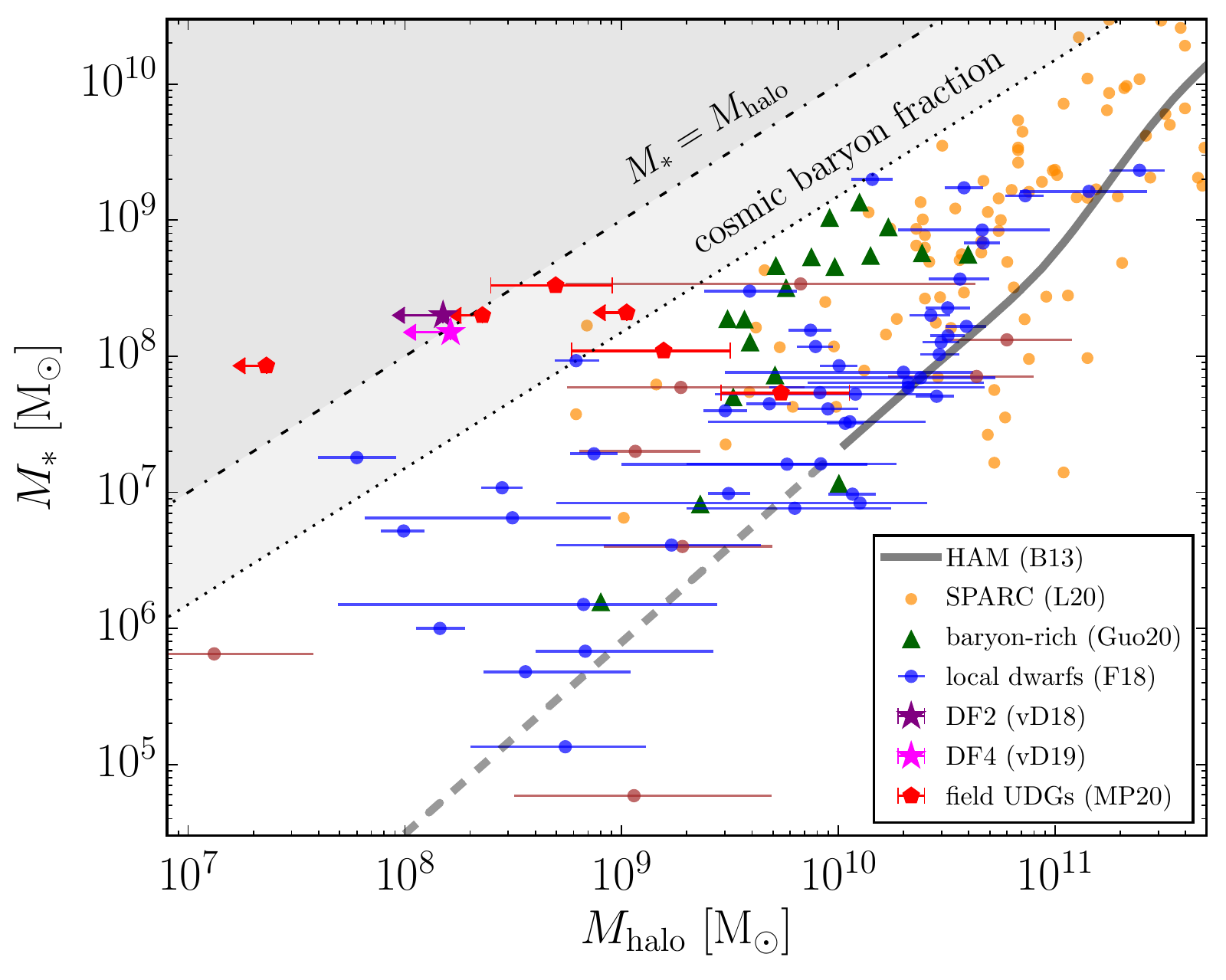}
    \caption{Stellar mass of galaxies as a function of their DM halo mass at $z=0$. The solid line corresponds to the mean relation obtained from HAM and other observational constraints by \citet{Behroozi13}. The points show results of dynamical modelling of predominantly isolated nearby galaxies from \citet{Forbes18b}, \citet{Li20}, and the triangles show the baryon-dominated galaxies from \citet{Guo20}. The \citet{Forbes18b} data is coloured according to morphology (blue: late types; brown: early types). The pentagons are simple mass models based on $\HI$ rotation curves from \citet{ManceraPina20}, and the stars are dynamical inferences for DF2 \citep{vanDokkum18a} and DF4 \citep{vanDokkum19}. For all the UDGs, the data probe out to $r\sim 8-10\kpc$, where DM is expected to dominate. The dotted line shows the upper limit set by the cosmological baryon fraction, and the dash-dotted line marks the region where $\Mstar\geq\Mhalo$. Error bars on the \citet{Li20} data are omitted for clarity. Taken at face value, the dynamical measurements indicate a large scatter around the mean relation. One of the key assumptions of our model is that the scatter at fixed halo mass is only limited by the cosmological baryon fraction. Note that most of the low-mass galaxies shown here (except DF2 and DF4 and the early types from \citet{Forbes18b}) contain significant amounts of \HI, which would shift their total baryon mass closer to or even above the cosmic baryon budget.}
\label{fig:SMHM}
\end{figure*}

Figure~\ref{fig:SMHM} shows the mean SMHM relation assumed in this work and compares it to a comprehensive set of dynamical models of nearby low-mass galaxies from \citep{Read17}, \citet{Forbes18b}, and the SPARC database \citep{Li20}. With the exception of a few Local Group members, most objects are isolated late-type galaxies\footnote{Galaxies in high density environments could have their DM halo masses reduced through tidal stripping, artificially increasing the intrinsic scatter in the SMHM relation. However, while 18 of the \citet{Forbes18b} late-types are not isolated, their high $\HI$ content indicates a lack of significant stripping of their DM haloes.}. The mass models use a variety of different techniques, but all of them account for the possibility of a cored DM halo using coreNFW DM density profile \citep{Read16b}. For comparison, the figure also includes simple dynamical estimates based on $\HI$ observations of gas-rich UDGs by \citet{ManceraPina20}, baryon-dominated galaxies \citep{Guo20}, and velocity dispersion measurements for NGC1052-DF2 and DF4 \citep{vanDokkum18a,vanDokkum19}. We refer the reader to Appendix~\ref{sec:massmodels} for details on each of these datasets and mass models. In general, dynamical measurements have significant uncertainties and may suffer from systematics which are difficult to estimate. However, across several datasets, these dynamical models suggest the possibility of large amounts of scatter, in excess of the uncertainties, around the mean SMHM relation obtained from abundance matching\footnote{Large scatter in $\Mstar$ at fixed $\Mhalo$ would bias the estimation of the mean $\Mhalo(\Mstar)$ relation due to the steeply increasing number of galaxies at low masses that scatter upwards in stellar mass. However, in our analysis we do not estimate population statistics using observations, and therefore do not need to include this effect.}. The cosmic baryon fraction $\fbar \equiv \Omegabar/\Omegamatter \approx 0.15$ \citep{Planck15} places a strict upper limit on the galaxy stellar (and total baryonic) mass for a given halo mass. However, many galaxies appear to exceed it. This is another way of stating that these galaxies appear to be ``DM-deficient'', as this feature is commonly described. In fact, except for NGC1052-DF2 and DF4 and a few early-types (brown points), most of the objects in these samples are \HI-rich (with median gas fractions in the range $\sim 2-5$ for $\Mstar<10^9\Msun$), and therefore their total baryon content is even closer to the cosmic mean \citep[see][for extreme examples]{ManceraPina20,Guo20}. 

In our model, galaxies are initialised on a uniform grid in the $\Mstar - \Mhalo$ plane at $z=0$. In the following analysis we focus on the region of the plane extending down to 1 dex below the mean SMHM, and up to the maximum stellar mass allowed by the cosmic baryon fraction.

\subsection{Relating stellar mass and DM halo formation time}

A key step in the model is the link between galaxy stellar mass and host DM halo formation time. We assume a simple deterministic linear relation between the scatter in $\Mstar$ at fixed $\Mhalo$, $\log({\Mstar/\Mstarmean})$, and the scatter in halo concentration at that mass, $\log c - \log\langle c \rangle$. The structure of dark matter haloes in cosmological simulations is well described by the two-parameter NFW density profile \citep{Navarro97},
\begin{equation}
    \rho(r) = \frac{4\rhos}{r/\rs \left(1 + r/\rs \right)^2} ,
\end{equation}
where 
\begin{equation}
    \rs \equiv \R200/c
\end{equation}
is the scale radius that marks the transition in the logarithmic slope from $-1$ to $-3$, $\rhos$ is the density at the scale radius, and $c$ is the halo concentration. The halo virial radius is
\begin{equation}
    \R200 \equiv \frac{3\Mhalo}{4\pi (200\rhocrit)} ,
\label{eq:r200}
\end{equation}
where $\rhocrit$ is the critical density of the Universe. The density at the scale radius is related to the concentration parameter through the equation
\begin{equation}
    \rhos = \frac{200\rhocrit}{12} \frac{c^3}{\left[ \ln(1+c) - c/(1+c)\right]} .
\end{equation}
Cosmological DM-only simulations find an anti-correlation between halo mass and concentration \citep{DuttonMaccio14},
\begin{equation}
    \log c(\Mhalo) = 0.905 - 0.101\log \left(\frac{\Mhalo}{10^{12}\Msunh}\right) ,
\end{equation}
such that low-mass haloes have higher concentrations (and therefore higher central densities) than higher mass haloes. The scatter in concentration at fixed $\Mhalo$ is found to have a constant logarithmic width $\sigmac = 0.11$ \citep{DuttonMaccio14}. In the \LCDM~cosmology, the concentration is set by the formation time of the DM halo \citep{wechsleretal02},
\begin{equation}
    c = \frac{c_1}{\acoll} ,
\end{equation}
where $\acoll = (1+\zcoll)^{-1}$ is the scale factor at the time when the logarithmic mass accretion rate $\diff\log M/\diff\log a$ drops below 2, and $c_1 = 4.1$. This implies that low-mass haloes formed earlier than higher mass haloes, and that at a fixed mass, haloes with higher central densities (i.e. higher concentration) formed earlier than lower density haloes.

The maximum circular velocity,
\begin{equation}
    \Vmax \equiv \sqrt{\frac{{\rm G}M(<r)}{r}} ~\Bigg|_{\rm max} ,
\end{equation}
is a proxy for the central gravitational potential, and increases with both halo mass and concentration, such that haloes with higher concentration have higher $\Vmax$ at fixed $\Mhalo$.

In HAM models, observed galaxy clustering is best reproduced when assuming a (positive) correlation between galaxy stellar mass and halo maximum circular velocity $\Vmax$ \citep{Campbell18}. Based on physical arguments, $\Vmax$, which directly traces the central gravitational potential, is expected to be the halo property that sets the gas accretion rate and therefore the baryonic budget of galaxies. This implies a correlation between halo assembly time and $\Mstar$ at fixed $\Mhalo$, which is also observed in cosmological hydrodynamical simulations tuned to reproduce the overall properties of galaxies at $z=0$. In the EAGLE simulations \citep{Schaye15,Crain15}, galaxies that scatter upwards from the mean SMHM relation are hosted by earlier-forming DM haloes because these were able to retain a larger fraction of their baryons (due to their deeper potential) and had a longer time to grow their stellar mass \citep{Matthee17,Kulier19}. The same correlation between halo concentration and $z=0$ stellar mass is also found in the Illustris \citep{Artale18} and IllustrisTNG simulations \citep{Bose19}, and in the EMERGE semi-empirical model \citep{Moster20}. Following this argument, we assume a simple linear scaling between concentration and stellar mass at fixed halo mass,
\begin{equation}
    \log c = \log\langle c \rangle + \fc\sigmac \log(\Mstar/\Mstarmean) ,
\end{equation}
where the free parameter $\fc=0.5$ controls the slope of the relation. The relation implies that galaxies that inhabit haloes with concentrations 1$\sigma$ away from the mean (at a given $\Mhalo$) should have stellar masses that deviate from the mean SMHM by 2 orders of magnitude. This large scatter in $\Mstar$ assumes that galaxies with such a large deviation in stellar masses are relatively common in haloes with $\Mhalo \la 10^{10}\Msun$. This precise form of this relation is difficult to deduce from observations because of the effect of SN feedback on the halo mass profiles. The value of $\fc$ is observationally unknown in low mass galaxies, and the value we use here is arbitrary and corresponds to a shallow correlation between concentration and stellar mass (at fixed $\Mhalo$). This choice produces good agreement with UDG mass profiles (see Section~\ref{sec:UDGs}), and its value does not impact the predictions for galaxy structure significantly (as shown in Appendix~\ref{sec:parameters}). As shown in Figure~\ref{fig:SMHM}, this is consistent with what is found in inferences from dynamical modelling of dwarf galaxies with $\Mstar \la 10^8\Msun$ \citep{Ferrero12,Schneider17,Forbes18b,Li20}. The sensitivity of the predictions to the value of $\fc$ is explored in Appendix~\ref{sec:parameters}.

\begin{figure*}
    \includegraphics[width=0.8\textwidth]{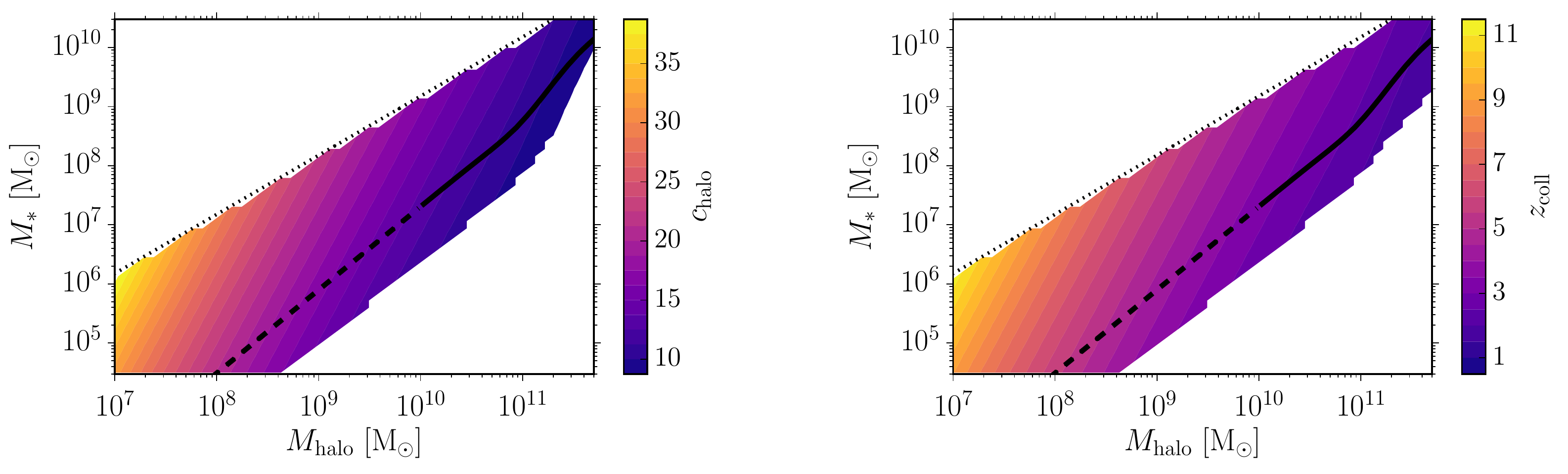}
    \caption{Predicted host DM halo properties as a function of the position of the galaxy in the $\Mstar-\Mhalo$ plane. Left: DM halo concentration at $z=0$. Right: collapse redshift. The dotted line shows the cosmic baryon fraction. At a fixed halo mass, galaxies above the mean SMHM relation are hosted by haloes with higher concentrations and earlier collapse times.}  
\label{fig:model1}
\end{figure*}

The left and right panels of Figure~\ref{fig:model1} show the DM halo concentration and collapse redshift, respectively, on the plane defined by $\Mstar$ and $\Mhalo$. Galaxies with $\Mhalo \sim 10^{10}\Msun$ that lie on the mean SMHM relation have concentrations $\sim 13$, while galaxies that scatter maximally above the relation, with $\Mstar = 1.5\times10^9\Msun$, have concentrations as large as $\sim 17$. At this halo mass, this means that the collapse redshift of a galaxy on the SMHM relation is $\zcoll \sim 2$, while it can reach $\zcoll \gtrsim 3$ above it. 

To summarise, at a fixed halo mass, galaxies above the mean SMHM relation have higher concentrations, and earlier collapse times.

\subsection{Star-forming environment at the collapse epoch}
\label{sec:environment}

Next, we need to determine the star-forming environment at the collapse epoch of the halo. First, in order to assign sizes to galaxies based on their halo mass, we use the empirical relation from \citet{Kravtsov13},
\begin{equation}
    \re = 0.015 \R200 ,
    \label{eq:r_e_z0}
\end{equation}
which relates the effective radius $\re$ of the galaxy to the virial radius $\R200$ of its DM halo at $z=0$. This relation is found to hold over 8 orders of magnitude in stellar mass and across all morphological types \citep{Kravtsov13}. Given that the relation was calibrated using halo abundance matching, we assume that galaxy sizes depend only on $\Mhalo$ through equations~(\ref{eq:r200}) and (\ref{eq:r_e_z0}). Since the \citet{Kravtsov13} relation only evolves by $\sim 50$ per cent between $z=3$ and $z=0$ \citep{Mowla19}, we make the simplifying assumption that it is constant in time. The effective radius can be obtained at any epoch using the predicted halo mass growth histories implemented in the {\sc commah} software package \citep{Correa15a, Correa15b, Correa15c},
\begin{equation}
    \re^{\zcoll} = 0.015 \R200^{\zcoll} .
    \label{eq:r_e}
\end{equation}
As discussed above, the aim of this model is to investigate the effect of feedback from stellar clusters on the structure of their host galaxies. Models that follow the formation and evolution of GCs in cosmological simulations find that earlier-forming DM haloes host a larger number of GCs at $z=0$ \citep[][Table B1]{emosaicsII}. This suggests that the high gas accretion rates during halo collapse drive the formation of a significant fraction of GCs. We therefore assume the peak GC formation epoch is the collapse epoch of the halo \citep[also see e.g.][]{Reina-Campos19}. We estimate the galaxy SFR at this epoch using the time-averaged SFR, $\Msdot$, at $z > \zcoll$,
\begin{equation}
    \SFR = \frac{\Mstar^{\zcoll}}{\tcoll} ,
    \label{eq:SFR}
\end{equation}
where $\tcoll$ is the halo collapse time (i.e. the time since the Big Bang at $\zcoll$), and
\begin{equation}
    \Mstar^{\zcoll} = \fstar \Mstar 
    \label{eq:Mstar}
\end{equation}
is the stellar mass at $z=\zcoll$. The parameter $\fstar$ sets the fraction of the total stellar mass formed before $\zcoll$, and is set to $\fstar=0.2$ in the fiducial model. This value is broadly consistent with the star formation histories of low-mass galaxies in semi-empirical models \citep[e.g.][]{Moster13,Behroozi13}. In Appendix~\ref{sec:parameters} we examine the sensitivity of the results to this assumption. The SFR surface density then follows from the SFR and the effective radius of the galaxy,
\begin{equation}
    \SigmaSFR = \frac{0.5~\SFR}{\pi (r_e^{\zcoll})^2} ,
    \label{eq:SigmaSFR}
\end{equation}
where the factor 0.5 accounts for the assumption that the SFR follows the distribution of the stellar mass, where half is within $\re$. We use the Kennicutt-Schmidt relation measured at $z\sim 1{-}3$ by \citet{Genzel10} to obtain the mean gas surface density within the effective radius from $\SigmaSFR$. This relation is slightly shallower than the original \citet{Kennicutt98} fit, but also provides a good fit to the $z=0$ data. 

After calculating the mean gas surface density within $\re$, we then assume (following \citealt{DiCintio16}) that the cold gas distribution is described by an exponential profile with a scale length $\sim 2$ times larger than for the stellar component, 
\begin{equation}
    \regas^{\zcoll} = 2\re^{\zcoll} .
\end{equation}
This allows us to calculate the total gas mass,
\begin{equation}
    \Mgas^{\zcoll} = 2\pi (\re^{\zcoll})^2 \Sigmagas,
\end{equation}
and the gas fraction at the collapse epoch,
\begin{equation}
    \fgas = \frac{\Mgas^{\zcoll}}{\Mstar^{\zcoll} + \Mgas^{\zcoll}} .
\end{equation}
Lastly, the DM density profile, together with the stellar and gas mass profiles at the collapse epoch can be used to obtain the total mass enclosed within the disc effective radius, and hence the rotation frequency of the disc,
\begin{equation}
    \Omega = \sqrt{\frac{G[\MDM^{\zcoll}(<\re) + \Mstar^{\zcoll}(<\re) + \Mgas^{\zcoll}(<\re)]}{\re^3}} .
\end{equation}
 
\begin{figure*}
    \includegraphics[width=0.8\textwidth]{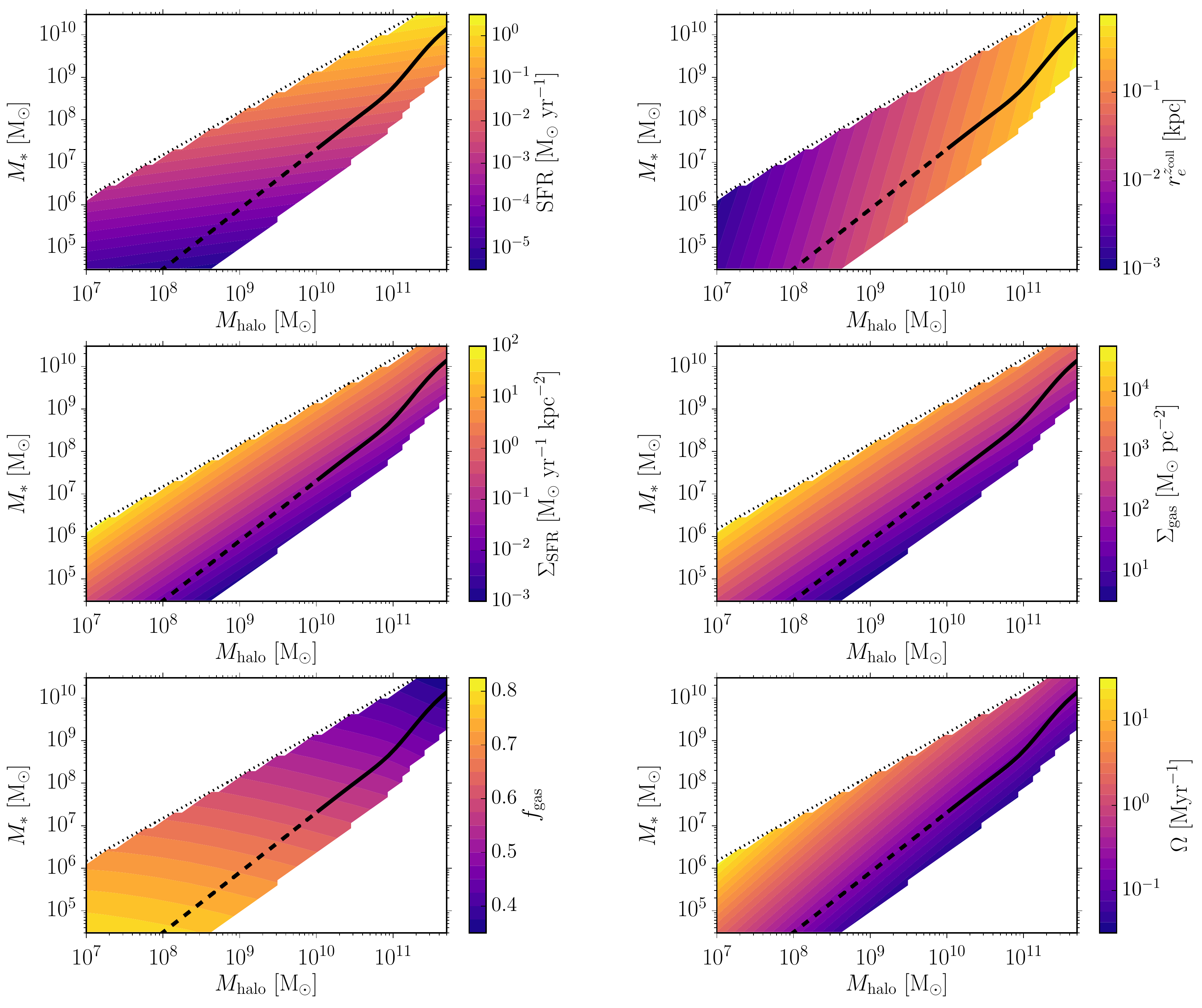}
    \caption{Dependence of galaxy properties \emph{at the collapse redshift} on the present position in the $\Mstar-\Mhalo$ plane. First row: time-averaged star formation rate and galaxy effective radius at $z=\zcoll$. Second row: SFR surface density (left) and gas surface density (right). Last row: gas fraction (left) and disc rotation frequency (right). The solid line is the mean SMHM relation from \citet{Behroozi13}, and the dotted line indicates the cosmic baryon fraction. At a fixed halo mass, galaxies that scatter above the mean SMHM relation had increasingly higher SFRs, and smaller sizes. This resulted in higher mean SFR surface densities, mean gas surface densities, global gas fractions, and rotation frequencies at their formation epoch, $z=\zcoll$.} 
\label{fig:model2}
\end{figure*}

Figure~\ref{fig:model2} shows the predictions for the SFR, $\re$, $\SigmaSFR$, $\Sigmagas$, $\fgas$, and $\Omega$ at the collapse time as a function of $\Mstar$ and $\Mhalo$. At a fixed halo mass of $\Mhalo=10^{10}\Msun$, the SFR increases from $\sim 2\times10^{-3}\Msunyr$ at the mean $\Mstar$, to $\sim 0.2\Msunyr$ for the maximum upward scatter. Disc effective radii (at the collapse time) decrease only slightly from $\sim 0.09\kpc$ on the mean relation to $\sim 0.06\kpc$ above it. As expected from the trend in the SFR and galaxy sizes, the SFR surface density can increase by a factor of $\sim 200$ between the  mean and the maximum stellar mass at this halo mass. The gas surface density also increases by a factor of $\gtrsim 100$. The gas fraction decreases by about 10 per cent, and the angular frequency at $\re$, which is a proxy for the shear, increases substantially by up to a factor of $\sim 15$. Upwards scatter from the mean SMHM relation therefore results in galaxies with larger stellar and gas surface densities, and hence larger gas pressures. As we will show in the next section, these conditions lead to increased efficiency of star cluster formation.    

\subsection{Forming stellar cluster populations}
\label{sec:clustering}
 
Stellar cluster populations can be described at birth in terms of the fraction of star formation occurring in bound clusters, and their initial mass distribution (i.e. the ICMF). 

Under the hypothesis of a differentially-rotating disc in hydrostatic equilibrium, these quantities can be determined from the gas surface density, the angular rotation frequency, and the Toomre $Q$ parameter \citep{Kruijssen12d,Trujillo-Gomez19}. We assume the value $Q=0.5$, which is typically observed at $r=\re$ in $z\sim 2$ star-forming galaxies \citep{Genzel14}. We calculate the bound fraction $\fbound$ with the model presented in \citet{Kruijssen12d}, and assume that the ICMF is well described by a double Schechter function \citep{Schechter76} with low and high-mass truncation masses $\Mmin$ and $\Mmax$, respectively \citep{Trujillo-Gomez19}. In the ICMF model, the low-mass truncation is set by the mass scale within the hierarchical cloud structure that can remain bound after gas expulsion by stellar feedback. The high-mass truncation is set by the fraction of the largest shear-limited gas cloud that is able to collapse and form stars before feedback disrupts it.

These models generally predict a steep increase of the bound fraction and both the minimum and maximum cluster masses with increasing gas surface density. An additional, shallower dependence on $\Omega$ reduces $\Mmax$ as the degree of shear support increases \citep[for details, see][]{Reina-Campos17, Trujillo-Gomez19}. These environmental models reproduce the  observed populations of young clusters in the nearby Universe, finding good agreement with the observed cluster formation efficiency \citep[e.g.][]{Ginsburg18,Adamo20a,Adamo20b}, and cluster mass functions \citep{Messa18,Trujillo-Gomez19} across a wide range of galactic environments \citep[also see][]{Pfeffer19}. The bound fraction and maximum cluster mass models have been implemented in cosmological galaxy formation simulations, and are found to produce GC populations with mass and metallicity distributions, and specific frequencies in good agreement with observations \citep{Pfeffer18,emosaicsII}.
 
Together, the stellar mass of the galaxy, the bound fraction, and the ICMF completely determine the number and masses of the star clusters formed at $z \sim \zcoll$. The ICMF is given by
\begin{equation}
    \frac{{\rm d}N}{{\rm d}M} = \Phi_{\rm norm} M^{-2} \exp\left( -\frac{\Mmin}{M} \right) \exp\left( -\frac{M}{\Mmax} \right) ,
    \label{eq:ICMF}
\end{equation}
where the normalisation factor $\Phi_{\rm norm}$ is obtained by requiring that the total mass under the ICMF equal the fraction of the galaxy stellar mass in bound clusters, 
\begin{equation}
    \Mcl = \fbound \Mstar^{\zcoll} .
\end{equation} 
To compute the number of star clusters, we neglect the effects of random sampling and simply integrate the ICMF,
\begin{equation}
    \Ncl = \int_{0}^{\infty}{ \frac{\diff N}{\diff M} \diff M} .
\end{equation}
We define GCs as star clusters of initial mass $M>10^5\Msun$, and obtain their total number by integration of the ICMF: 
\begin{equation}
    \NGC = \int_{M_{\rm GC, min}}^{\infty}{ \frac{\diff N}{\diff M} \diff M} ,
\end{equation} 
where $M_{\rm GC, min} = 10^5\Msun$. This mass is near the observed median GC mass across most galaxies \citep{Jordan07}, and also corresponds to the minimum mass of clusters that are expected to survive disruption for a Hubble time in low-mass galaxies \citep{Kruijssen15b}.
We explore the effect of varying this threshold mass in Section~\ref{sec:global}. Lastly, the specific frequency is a measure of the number of GCs per unit galaxy stellar mass, and it is defined as 
\begin{equation}
    \TN = \frac{\NGC}{\left( \Mstar/10^9\Msun \right)} .
    \label{eq:T_N}
\end{equation}
 
\begin{figure*}
    \includegraphics[width=0.8\textwidth]{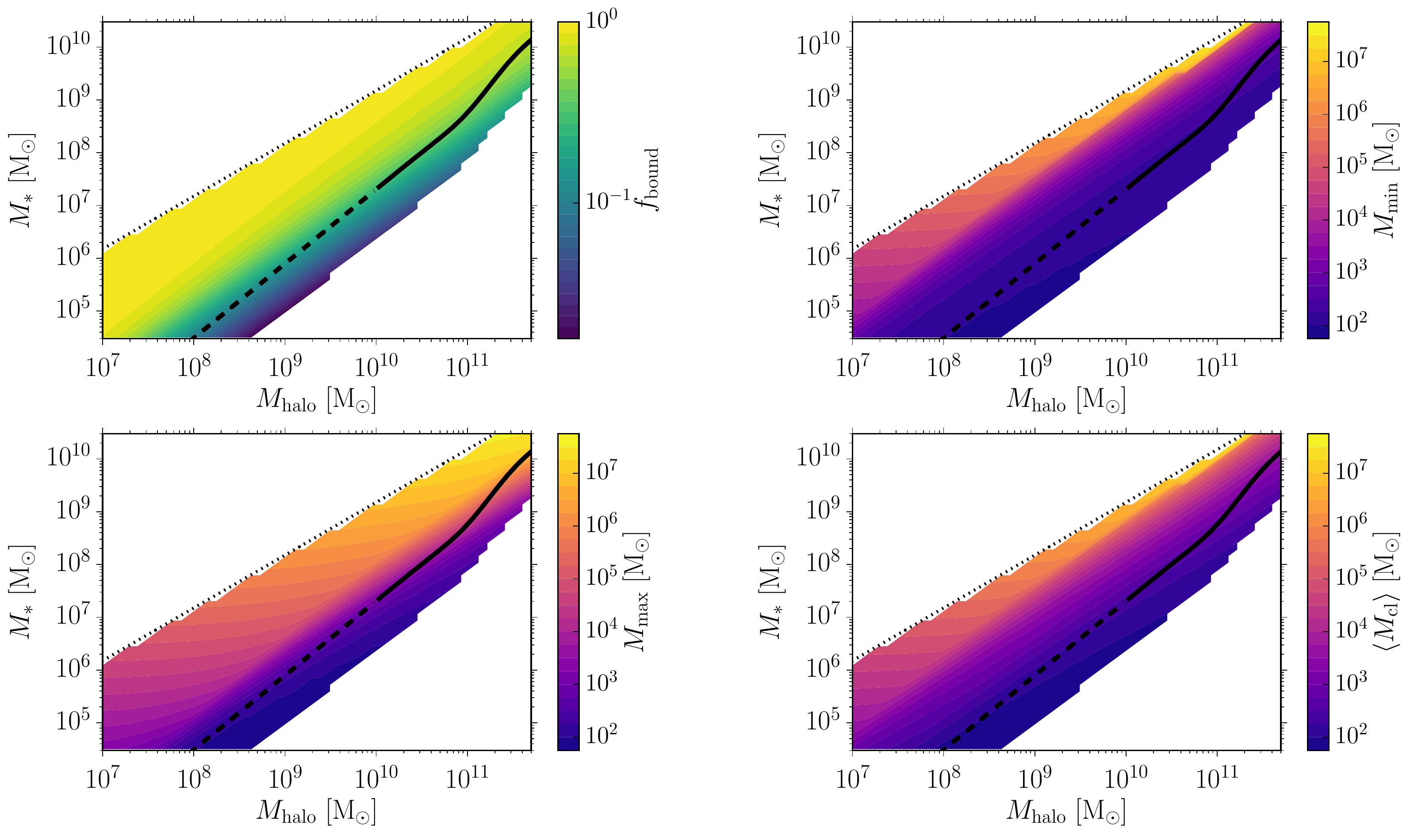}
    \caption{Demographics of star cluster populations formed \emph{at the collapse epoch} as a function of the present position of the galaxy in the $\Mstar-\Mhalo$ plane. First row: gravitationally bound fraction of star formation (left), and minimum cluster mass (right). Second row: maximum cluster mass (left), and mean cluster mass (right). The solid line is the mean SMHM relation from \citet{Behroozi13}, and the dotted line indicates the cosmic baryon fraction. At a fixed halo mass, galaxies that scatter above the mean SMHM relation show a steep increase in the fraction of star formed in bound clusters, accompanied by a rapid increase in $\Mmax$ and a slower increase $\Mmin$. Together, the increase in both $\Mmin$ and $\Mmax$ drives a steep increase in the mean cluster mass. Galaxies near the cosmic baryon fraction will form a higher fraction of stars in clusters, and more massive cluster populations at $z=\zcoll$ compared to galaxies on the SMHM relation. }  
\label{fig:model3}
\end{figure*}

Figure~\ref{fig:model3} shows the effect that the departure from the mean SMHM relation has on the cluster populations formed at the collapse epoch of the halo. Driven by the steep increase in gas pressure (traced by $\Sigmagas$), $\fbound$ can increase from $\sim 0.3$ to $1.0$ for galaxies with $\Mhalo=10^{10}\Msun$ as they scatter upwards from the relation\footnote{These values are relatively high compared to nearby galaxies, and originate from the high values of $\Sigmagas$ at $\zcoll$. The corresponding gas surface densities are comparable to observed massive star-forming galaxies at $z>1$ \citep{Tacconi13}}. The minimum cluster mass follows a similar trend. For $\Mhalo=10^{10}\Msun$, galaxies on the relation form clusters with $\Mmin \sim 10^2\Msun$, which increases to $\gtrsim 10^3\Msun$ for an upward scatter of $\sim 1$~dex, and then steeply increases to $\gtrsim 10^6\Msun$ near the cosmic baryon fraction. The maximum truncation mass also increases as galaxies scatter upwards from the relation\footnote{The model predicts no massive clusters in dwarfs similar to Fornax, which lies near the mean SMHM relation with $\Mhalo\approx10^{10}\Msun$, and contains 5 GCs with $M\sim10^5\Msun$. However, at this mass, galaxies hosting such large GC populations seem to be rare \citep{Shao20}. A single gas-rich major merger (as opposed to many minor mergers) could in principle account for the majority of Fornax's GCs \citep{Trujillo-Gomez20b}. Appendix~\ref{sec:toomre} shows that assuming a larger value of Toomre $Q=2.0$ can account for the GC population of Fornax without significantly altering the model predictions for UDGs.}, with a secondary trend of increase towards more massive DM haloes due to their larger discs producing weaker shear (see Figure~\ref{fig:model2}). Together, these trends result in the steep growth of the population of massive clusters with increasing $\Mstar$ at fixed $\Mhalo$. The mean cluster mass also increases steeply with upwards departure from the SMHM relation, reaching $\ga 10^6\Msun$ in galaxies with stellar masses near the cosmic baryon fraction. The increase in $\fbound$ and $\Mclmean$ for galaxies above the SMHM relation implies that they produce a larger fraction of stars in clusters, and that these clusters are significantly more massive than for galaxies on the relation. Section~\ref{sec:global} shows how this impacts GC populations, and in the next subsection, we explore how these GC populations impact the evolution of galaxy structure. 

\subsection{Feedback-driven expansion}
 
Having defined the gas, stellar, and star cluster content of the galaxy at the collapse epoch of the halo, we can now estimate the effect of supernova feedback from both clusters and field stars on the structural evolution of the galaxy from $z=\zcoll$ until $z=0$. For this, we assume that massive gas outflows can generate large enough fluctuations in the gravitational potential to irreversibly heat the orbits of DM particles and reduce their central density, forming a shallow core in the halo density profile. This assumption is supported by a vast body of literature \citep[e.g.,][]{Navarro96b,ReadGilmore05,Mashchenko08,PontzenGovernato12,Teyssier13,DiCintio14,Onorbe15,Read16a,Freundlich20,Burger21}. Two elements are necessary to predict the effect of feedback from clustered star formation in our model: the impact of SN clustering on the energy loading of galactic winds, and the coupling of the wind energy to the gravitational potential energy of the DM halo. 

The first element is the fraction of total SN energy that drives galactic winds, and its dependence on the degree of spatial and temporal clustering of star formation. This has been studied extensively in analytical models and controlled numerical experiments where SNe are detonated uniformly or in a clustered (in space and time) fashion within a patch of the ISM \citep[e.g.,][]{Sharma14,Keller14,KimOstriker15,WalchNaab15,Girichidis16,Gentry17,Fielding18,Gentry19} or in entire galaxies \citep{Keller20b,Keller21}. \citet{Fielding17} investigated this effect on simulated galactic discs as a function of the number of overlapping SNe, the gas surface density, and the gas scale-height. They found that the resulting wind energies are consistent with those expected from superbubbles that break out of the gas disc when their cooling radius is large enough to exceed the scale-height. The wind energy loading (i.e. the fraction of SN energy that goes into driving a wind) in the simulations can be parametrized using the expression\footnote{To obtain this expression we adjusted equation (2) of \citet{Fielding17} to account for the power-law index of the dependence on $\fcl$ shown in the bottom panel of their figure 4. Since their equation does not include a normalisation, we also obtain it using the fits in their figure 4.}
\begin{equation}
    \etaw(h,\Sigmagas,\fcl) = 10^{-2} \left(\frac{h}{0.03\kpc}\right)^{-1.5}   \left(\frac{\Sigmagas}{10\Msunpc2}\right)^{-1} \fcl^{1.05} ,
    \label{eq:eta_E}
\end{equation}
where $h$ is the scale-height of the gas, and $\fcl$ is a measure of the degree of SN clustering. It corresponds to the number of supernovae progenitors formed per $100\Msun$ of stars, relative to the case with spatially uniform star formation. We use this expression to calculate the wind energy loading. The scale-height is obtained using eqs.~(33) and (34) from \citet{krumholzmckee05}, 
\begin{equation}
    h = \frac{\Sigmagas}{2\rhogas} ,
\end{equation}
which assume a disc in hydrostatic equilibrium and a flat rotation curve. The gas mid-plane density is
\begin{equation}
    \rhogas = \frac{ \pi G \phiP \Sigmagas^2 }{ 2 \sigma^2 }  ,
\end{equation}
where $\sigma$ is the gas velocity dispersion. The dispersion can be estimated using eq. (35) in \citet{krumholzmckee05},
\begin{equation}
    \sigmadyn = \frac{ \pi G Q \Sigmagas } { \sqrt{2}\Omega } ,
\end{equation}
where the factor $\phiP=3$ accounts for the gravity due to the stars, and $Q=0.5$ as before. This dynamical estimate of the gas velocity dispersion is corrected for thermal support using the weighted mean of the atomic and molecular components,
\begin{equation}
    \sigma = \sqrt{ (1-\fH2) \sigmaHI^2 + \fH2 \sigmaH2^2 } ,
\end{equation}
where the velocity dispersion of gas phase $X$ is 
\begin{equation}
\sigma_{X}^2 = \sigmadyn^2 + c_{{\rm s, }X}^2 . 
\end{equation} 
The sound speeds of the two components are set to $\csHI = 5.0\kms$ (i.e. $T_{\rm HI}\approx 3000~\rm{K}$), and $\csH2=0.3\kms$ \citep{krumholzmckee05}. Lastly, the molecular gas fraction is given by eq.~(73) in \citet{krumholzmckee05},
\begin{equation}
     \fH2 = \left[ 1 + 0.025\left(\dfrac{\Sigmagas}{10^2\Msunpc2}\right)^{-2} \right]^{-1} .
\end{equation}
For the high-redshift conditions of interest here, $\fH2 \approx1$ such that the contribution of thermal pressure to the velocity dispersion is very small. At the collapse epoch, a galaxy with $\Mhalo=10^{10}\Msun$ and the maximum allowed stellar mass (i.e. the cosmic baryon fraction) is predicted to have a scale-height a factor of $\sim 2$ smaller than a galaxy on the SMHM relation.

We define $\fcl$ as the average number of SNe per cluster in the bound component of star formation,
\begin{equation}
    \fcl = \frac{\Mclmean}{\MSN} , 
\end{equation}
where $\Mclmean$ is the mean cluster mass, and $\MSN = 99\Msun$ is the total mass of stars in a simple stellar population that contains at least one SN progenitor (assuming a progenitor mass of $8\Msun$ for a \citealt{chabrier03} IMF). At a fixed SFR, a higher $\fcl$ implies an increase in both the spatial and in the temporal clustering of SNe because a larger fraction of SNe detonate simultaneously in massive star clusters. The total wind energy loading of the galaxy is then the weighted sum of the bound (i.e. star cluster) and unbound (i.e. field star) components,
\begin{equation}
    \etagal = (1.0 - \fbound)\etaw(h,\Sigmagas,1.0) + \fbound\etaw(h,\Sigmagas,\fcl) ,
\end{equation}
where the value $\fcl=1.0$ for the unbound component reflects the normalization of the SN rate surface density employed by \citet{Fielding17} in the case of uniformly distributed SNe. 
 
\begin{figure*}
    \includegraphics[width=\textwidth]{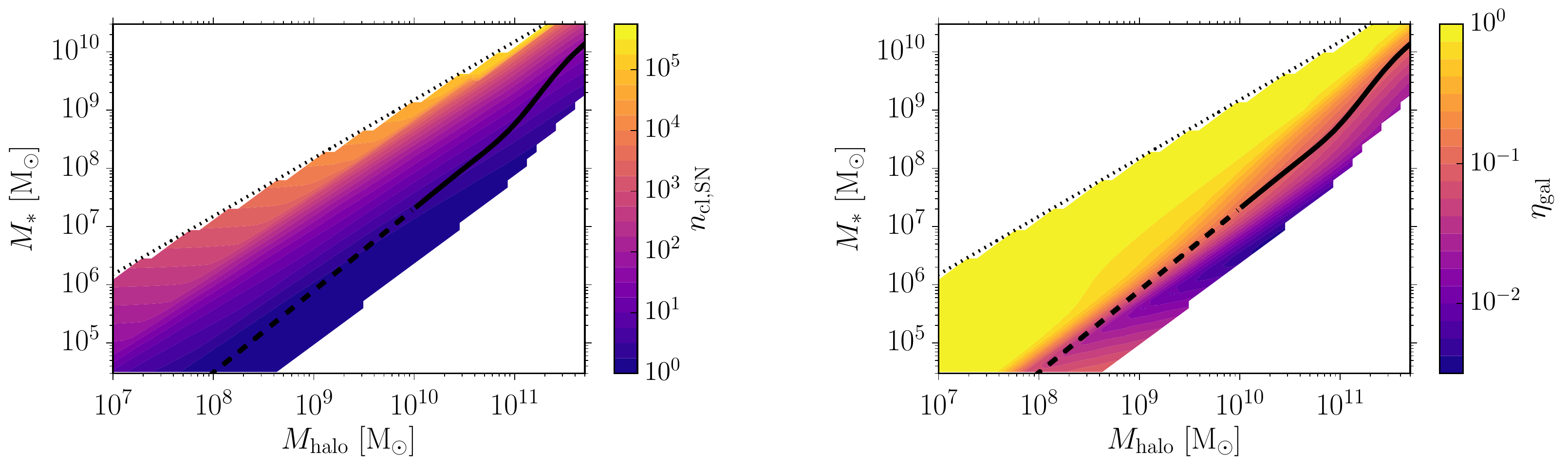}
    \caption{Impact of scatter away from the mean SMHM relation on the efficiency of galactic wind driving by supernovae. Left: Average SN clustering factor $\fcl$. Right: energy loading factor $\etagal$ of galactic winds due to SNe. The solid line is the mean SMHM relation from \citet{Behroozi13}, and the dotted line indicates the cosmic baryon fraction. The steep increase in the bound fraction of star formation and mean cluster mass as galaxies scatter upwards from the SMHM relation results in an increase of $\fcl$ of $\sim 3-5$ orders of magnitude. As stellar mass increases at a fixed halo mass, SN energy is more efficiently injected into the ISM to power galactic winds. This results in larger amounts of energy becoming available to dynamically heat the DM orbits and grow the DM core. Galaxies above the SMHM relation not only have larger SFR and therefore larger feedback budget, but are also more efficient at injecting feedback energy into galactic winds. } 
\label{fig:model4}
\end{figure*}

Figure~\ref{fig:model4} shows the impact of the degree of spatial and temporal clustering of star formation on the generation of galactic winds by SNe. The average number of overlapping SNe across the galaxy $\fcl$ systematically increases with the upward scatter in stellar mass at fixed halo mass (left panel), with a maximum near $\fcl \sim 10^3-10^{5.5}$ in galaxies near the cosmic baryon fraction. This is an increase of up to $\sim 3-5$ orders of magnitude relative to low-mass galaxies near the SMHM relation with $\fcl =1-10$. The right panel of Figure~\ref{fig:model4} shows the impact of stellar clustering on the energy loading of galactic winds at the collapse epoch. SNe become increasingly more efficient at powering galactic winds as their host galaxies scatter upwards from the SMHM relation. The fraction of SN energy driving the winds rises from $\etagal \la 0.1$ to a saturation at $\etagal=1$ as the scatter above the relation increases by $\gtrsim 1$ dex. This effect is driven by both the increase in SN clustering ($\fcl$), and the change in the underlying structure of the ISM (i.e. the decrease of the gas disc scale-height, see equation~\ref{eq:eta_E}). Note that the increase in $\fcl$ and $\etagal$ are both independent of the increase in the SFR of galaxies above the SMHM relation, and depend only on the structure of the cold ISM. In Section~\ref{sec:clustering_role} we show that the increased clustering of SNe in galaxies above the SMHM relation is the dominant factor driving the rise in the efficiency of galactic wind generation. 

Once the wind energy coupling is obtained, the second ingredient is the fraction of wind energy that produces potential fluctuations large enough to expand the DM distribution and reduce the central density of the halo. Unfortunately, there are no estimates of this value in the literature. However, values for the fraction of SN energy that couples to the DM cover a broad range $\sim 0.002{-}0.4$ \citep{Penarrubia12,Read16a}. \citet{Madau14} studied the time evolution of this coupling in cosmological simulations of dwarf galaxies, finding values $\sim 0.01{-}0.1$. In simulations with very different subgrid physics, \citet{Chan15} obtain total coupling fractions $\sim 0.005-0.1$. For simplicity, we assume that a constant fraction of the wind energy is transferred into the halo gravitational potential, 
\begin{equation}
    \etaDM = \fDM \etaw ,
\end{equation} 
and set the coupling to the upper bound obtained by \citet{Madau14}, $\fDM=0.1$. In Appendix~\ref{sec:parameters} we examine the dependence of the model predictions on this assumption. 

The complexity of the core formation process has prompted many numerical studies. There is general agreement that impulsive energy injection due to stellar feedback results in an overall flattening of the initially steep NFW density profile in the central few kiloparsecs \citep[see][for a review]{PontzenGovernato14}. \citet{Read16a} provide a convenient parametrization of the modified profile given by the coreNFW model,
\begin{equation}
    \McNFW(<r) = M_{\rm NFW}(<r) f^n ,
\end{equation}
where
\begin{equation}
    f^n = \left[ \tanh{\left(\dfrac{r}{\rcore}\right)} \right]^n ,
\end{equation}
and $n$ and $\rcore$ control the slope and size of the core, respectively. Their study shows that the DM haloes of isolated galaxy simulations are well fit when these parameters are given by
\begin{equation}
    n = \tanh{\left(0.04 \frac{\tSF}{\tdyn}\right)} ,
\end{equation}
and
\begin{equation}
    \rcore = 1.75 \re ,
    \label{eq:rcore}
\end{equation}
where $\tSF$ is the total duration of star formation in the galaxy, and $\tdyn$ is the circular orbit period at the scale radius of the NFW profile. This allows the DM core size $\rcore$ to be inferred from the observed effective radius of the stars $\re$. Because at the collapse redshift most galaxies have already been forming stars for several dynamical times, we assume $n=1$ throughout.

However, in our model we cannot obtain $\rcore$ by directly modelling the effect of stellar feedback on the galaxy effective radius. Instead, we use the DM coupling efficiency $\etaDM$ to obtain the size of the DM core. This is done by solving for the value of $\rcore$ that increases the gravitational potential energy by an amount equal to the total energy available from winds. This requires numerically solving the equation 
\begin{equation}
    \Delta W(\rcore) = \etaDM \NSN \ESN ,
\end{equation}
where $\NSN$ is the total number of SNe at $z>\zcoll$, $\ESN = 10^{51}\erg$ is the energy output of a single supernova, and
\begin{equation}
    \Delta W(\rcore) = -\dfrac{1}{2}  \int_{0}^{\infty}{ \dfrac{G [ M_{\rm fin}^2(\rcore) - M_{\rm init}^2 ] }{r^2}} \diff r
\end{equation}
is the difference in potential energy between the final coreNFW profile (and its embedded stellar disc), and the initial pristine NFW halo (and its expanded stellar disc) for a given value of $\rcore$. The initial and final total mass profiles are\footnote{Here we neglect the potential from the gas component because it is dissipative and couples directly to the SN energy.}
\begin{equation}
    M_{\rm init}(r) = M_{\rm NFW}(r) + \Mstar(\re^{\zcoll},r) ,
\end{equation}
and
\begin{equation}
    M_{\rm fin}(\rcore,r) = \McNFW(\rcore,r) + \Mstar(\re^{\rm fin},r) ,
\end{equation}
where $\Mstar(\re,r)$ is the stellar mass profile assuming an exponential stellar disc with effective radius $\re$. 

\begin{figure}
    \includegraphics[width=\columnwidth]{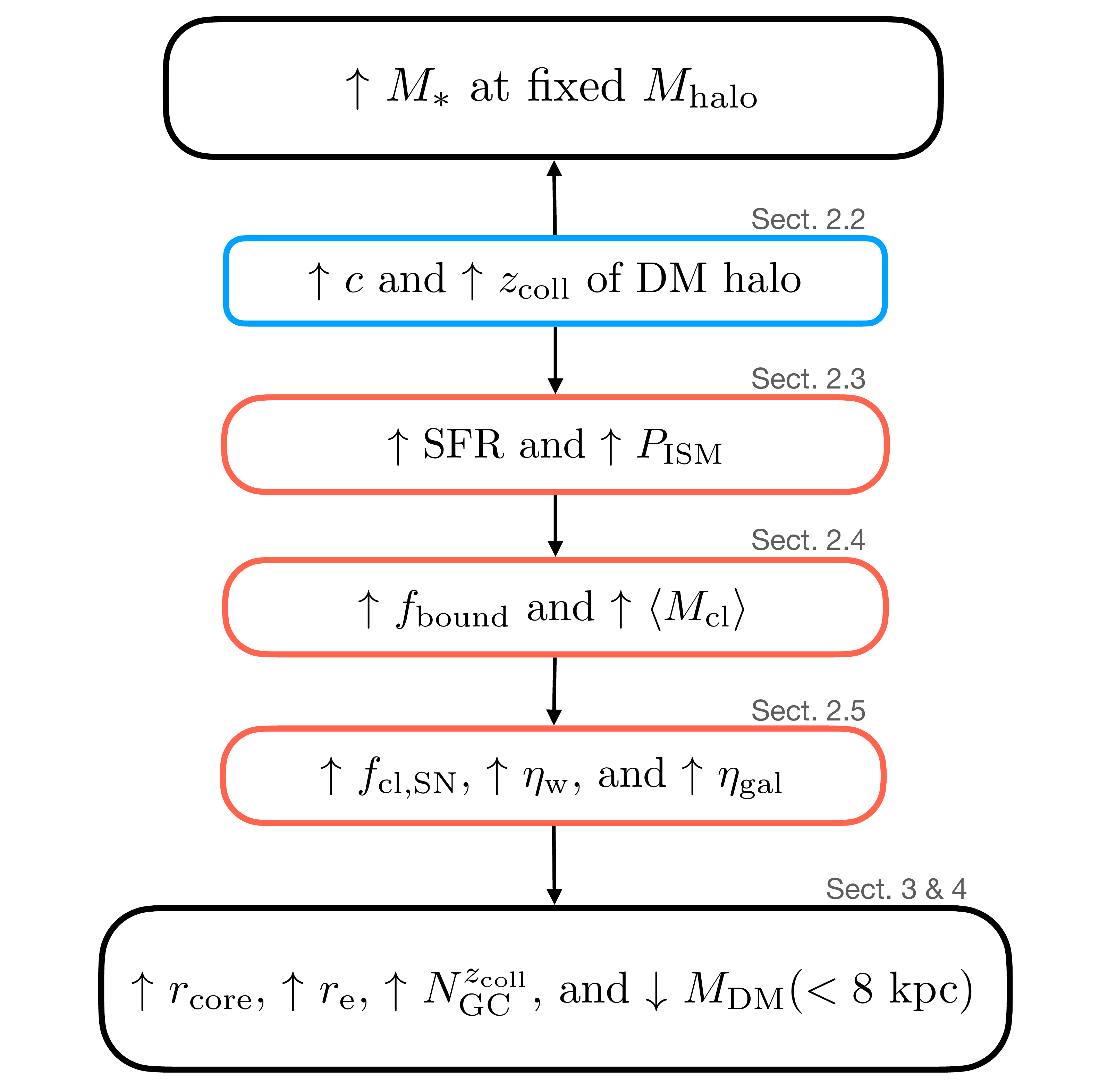}
    \caption{Schematic of the trends predicted by the model for a galaxy that scatters above the mean SMHM relation at $z=0$. The arrows indicate the direction of causal relations.}
    \label{fig:model_schematic}
\end{figure}

Lastly, we estimate the effect of expansion on the stellar disc. We denote as `old' the stellar disc already present at the collapse epoch that expands alongside the DM halo. The final $z=0$ effective radii of the `old' stellar disc is (Eq.~\ref{eq:rcore}),
\begin{equation}
    \reold = \re^{\rm fin} = \frac{\rcore}{1.75} .
\end{equation}
After the main episode of GC formation at $z=\zcoll$, we assume that the galaxy continues to form stars and grows in stellar mass, and that its feedback-driven winds have a gradually diminishing effect on the galaxy structure. The effective radius and stellar mass of the `young' disc that forms at $z<\zcoll$ are then given by (Eqs.~\ref{eq:r_e} and \ref{eq:Mstar})
\begin{equation}
    \reyoung = 0.015 \R200 ,
\end{equation}
and
\begin{equation}
    \Mstaryoung = (1-\fstar)\Mstar ,
\end{equation}
where the virial radius $\R200$ and stellar mass $\Mstar$ correspond to the values at $z=0$. The stellar mass distribution at $z=0$ is then determined by a smooth transition in the scale-length of the exponential disc between the old and the young populations,
\begin{equation}
    \re = \fstar \reold + (1-\fstar)\reyoung .
    \label{eq:r_e_total}
\end{equation}
This relation implicitly assumes a smooth evolution in time starting from the strong expansion of the old stellar component to the progressively weaker effect of expansion on increasingly younger populations. It also explicitly enforces the observed $z=0$ mass-size relation from \citet{Kravtsov13} for galaxies on the mean SMHM relation (see Section~\ref{sec:structure}). The quantities $\rcore$ and $\re$ are then completely specified by the $z=0$ position of a galaxy in the $\Mstar-\Mhalo$ plane, and define the mass distribution of the DM, and stellar components at present. For simplicity, we avoid the complication of predicting the gas mass and structure at $z=0$. Figure \ref{fig:model_schematic} provides a summary of the general trends predicted by the model.

\section{Effects of departure from the mean SMHM relation on galaxy evolution}
\label{sec:global}

In the previous section we introduced a semi-empirical model to describe the influence of clustered stellar feedback on galaxy structure. In this section we explore the demographics of the GC populations predicted by the model across the plane defined by $\Mstar$ and $\Mhalo$, and how these affect the evolution of the structure of their galaxies. 

\subsection{GC populations}

\begin{figure*}
    \includegraphics[width=\textwidth]{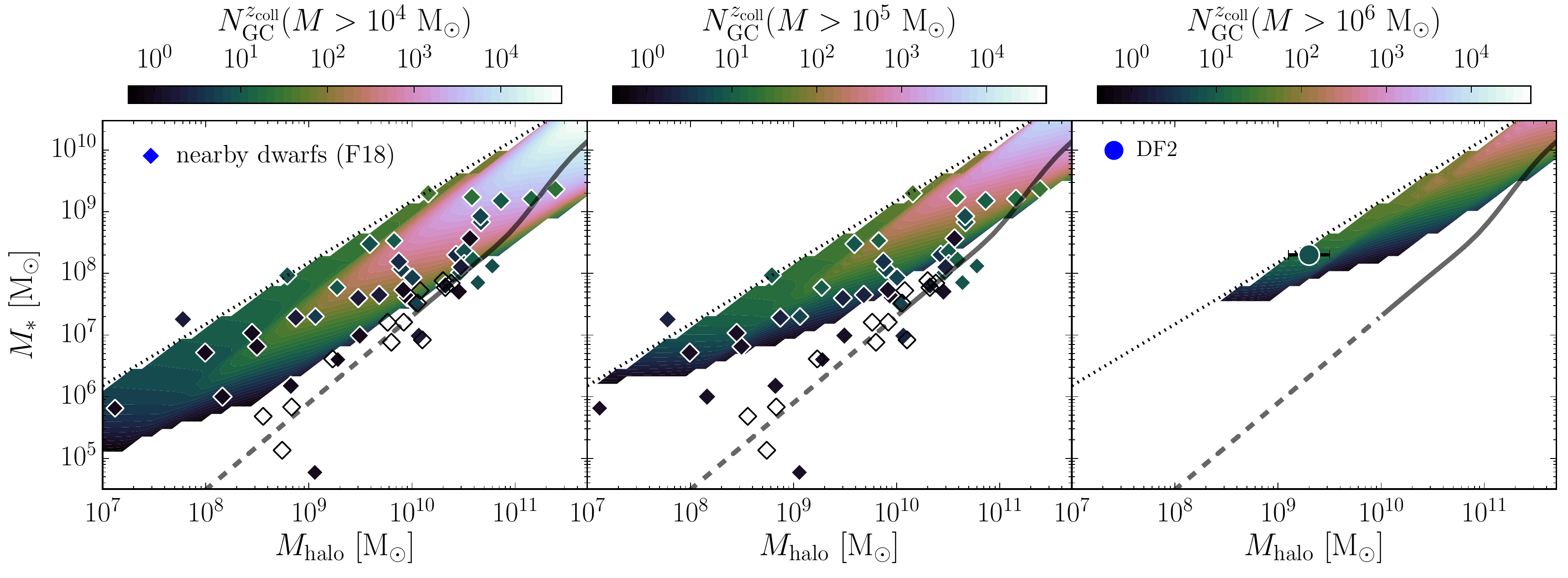}
    \caption{Demographics of the GC populations formed at $z=\zcoll$ as a function of the position of their host galaxy in the $\Mstar-\Mhalo$ plane. The panels show the total number of $M>10^4\Msun$ GCs (left), $M>10^5\Msun$ GCs (middle), and $M>10^6\Msun$ GCs (right). The sample of nearby dwarfs from \citet{Forbes18b} is shown as diamonds in the left and middle panels, with the colour of each symbol indicating the observed number of GCs hosted by each galaxy, and galaxies with no GCs indicated by empty symbols. The position of the galaxy NGC1052-DF2 is indicated by the circle with error bars, coloured according to the observed number GCs with $M>10^6\Msun$ (7 out of 11 in \citealt{vanDokkum18b}) (the error bars indicate the uncertainty in $\Mhalo$ from the range of best-fit halo masses obtained in Section~\ref{sec:DF2}). The solid line is the mean SMHM relation from \citet{Behroozi13}, and the dotted line indicates the cosmic baryon fraction. The model is in broad agreement with the trend found in the observations, in which galaxies with higher $\Mstar$ at fixed $\Mhalo$ host more GCs. Note that the \citet{Forbes18b} sample does not use a consistent lower mass limit to define a GC, and that low-mass dwarfs tend to host GCs of lower mass compared to massive galaxies.}
\label{fig:global_GCs1}
\end{figure*}

\begin{figure}
    \includegraphics[width=\columnwidth]{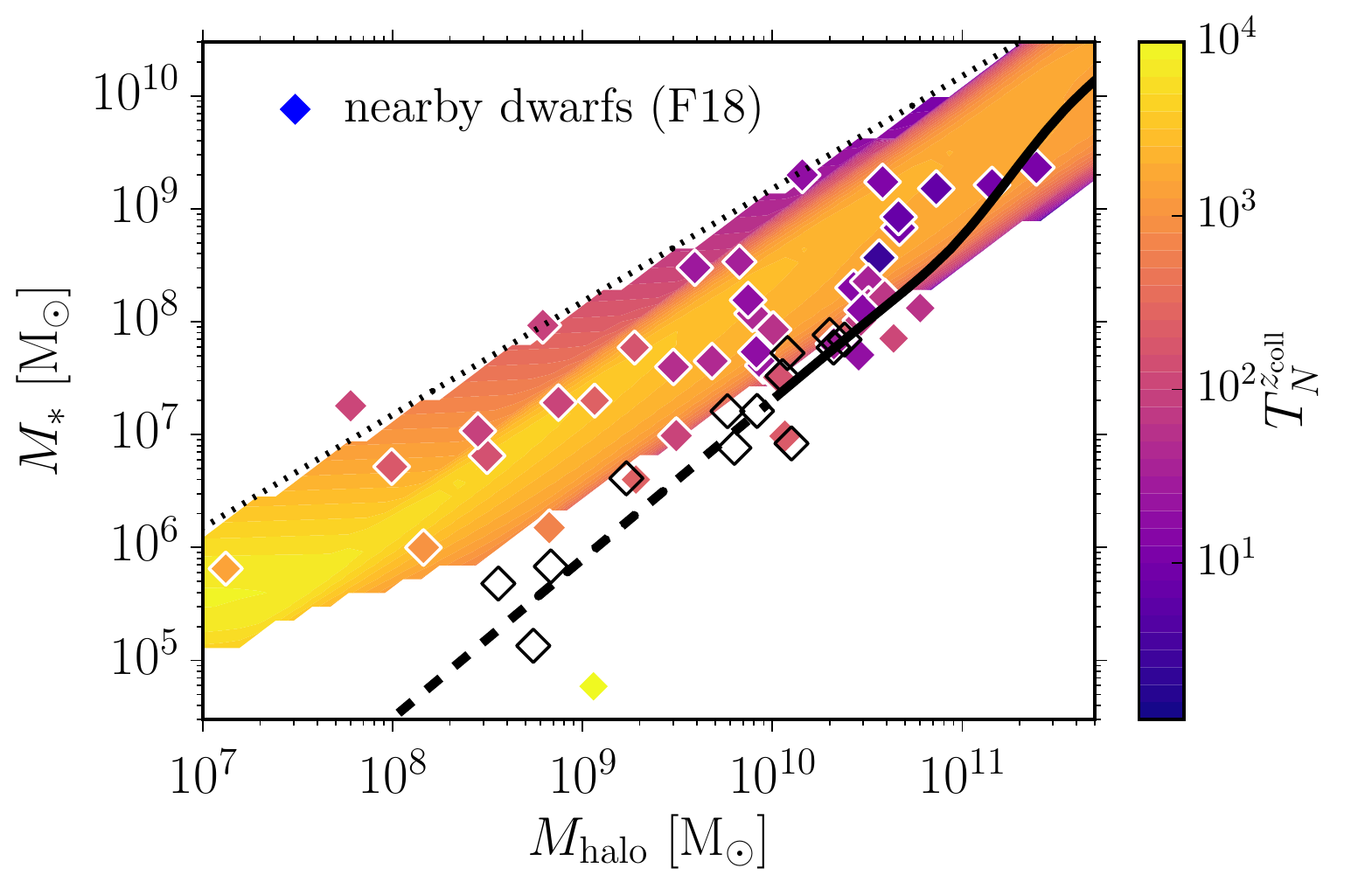}
    \caption{Abundance of galactic GC populations relative to stars at $z=0$ as a function of the position of their host galaxy in the $\Mstar-\Mhalo$ plane. The GC specific frequency quantifies the abundance of GCs relative to stars, $\TN \equiv \NGC (10^9\Msun/\Mstar)$, and is calculated here for clusters with $M>10^4\Msun$. The observational data for nearby low-mass galaxies from \citet{Forbes18b} is shown using diamonds coloured by $\TN$. Empty symbols correspond to galaxies that do not host any GCs. The solid line is the mean SMHM relation from \citet{Behroozi13}, and the dotted line indicates the cosmic baryon fraction. As seen in the observations, dwarf galaxies with lower stellar mass have more dominant GC populations relative to field stars. The model predicts an additional trend of increasing specific frequency with stellar mass at fixed halo mass which is also present in the \citet{Forbes18b} data. The drop in $\TN$ near the cosmic baryon fraction is due to a shift in the low-mass truncation of the ICMF. The predicted higher $\TN$ in galaxies that assembled earlier resembles what is observed in the Virgo cluster \citep{Peng08}.}
\label{fig:global_GCs2}
\end{figure}

Figures~\ref{fig:global_GCs1}, \ref{fig:global_GCs2}, and \ref{fig:global_structure} show the predictions of the model for the effect of scatter around the mean SMHM relation on the structure and GC populations of galaxies at $z=0$.

The total numbers of GCs (i.e. massive stellar clusters) with masses $M>10^4$, $M>10^5$, and $M>10^6\Msun$, are shown in Figure~\ref{fig:global_GCs1}. Upward scatter from the SMHM relation has a strong impact on the massive cluster populations of low-mass galaxies. This is caused by the earlier collapse times, the larger SFR surface densities (especially at early times), as well as the higher gas pressures of objects that lie above the relation. These conditions lead to higher fractions of bound clusters and larger mean cluster masses (see Section~\ref{sec:clustering} and Fig.~\ref{fig:model3}). The result is a steep increase in the number of massive GCs with increasing $\Mstar$ at a fixed $\Mhalo$. In addition, there is also an increase in the number of GCs $\NGC$ with increasing halo mass due to the larger $\Mmax$ in more massive haloes. In the model, galaxies with  $\Mhalo \la 10^{10}\Msun$ only form GCs when they scatter above the mean SMHM assumed here (see discussion in Section~\ref{sec:clustering}). This is consistent with the increasingly dominant role of mergers in the formation of GCs in low-mass galaxies in cosmological simulations \citep{Trujillo-Gomez20b}. The downturn in the number of GCs near the cosmic baryon fraction in the left and middle panels is due to the minimum cluster mass increasing past $10^4\Msun$, which shifts the ICMF towards more massive, but less numerous objects (see Figure~\ref{fig:model3}). The last panel of Figure~\ref{fig:global_GCs1} shows that galaxies with $\Mstar\gtrsim10^9\Msun$ located near the cosmic baryon fraction can form significant populations of very massive ($M>10^6\Mstar$) GCs.

For comparison, Figure~\ref{fig:global_GCs1} also shows the number of GCs hosted by each galaxy in the \citet{Forbes18b} sample of nearby dwarfs. These data are included in both the left and middle panels because the sample does not have a consistent minimum GC mass, and dwarf galaxies tend to contain more low-mass GCs than massive galaxies. Despite the limited sample size, the data suggest a trend of increasing number of GCs with stellar mass at fixed halo mass albeit with significant scatter. Since we do not model cluster dynamical mass loss, a quantitative comparison with the observations is not possible. However, the model qualitatively reproduces the general trend suggested by the data, predicting increasing GC numbers in galaxies with larger $\Mstar$ (at a fixed $\Mhalo$), although with a steeper dependence on $\Mstar$. Interestingly, observed dwarfs which do not host any GCs are located near the mean SMHM relation for $\Mhalo\la10^{10}\Msun$. This matches the region where no GCs are predicted to form. The right panel shows the good agreement between the number of massive GCs with $M>10^6\Msun$ found in DF2 by \citet{vanDokkum18b} and the model prediction. To place that galaxy in the $\Mstar$--$\Mhalo$ plane, we assume the best-fit halo mass obtained in Section~\ref{sec:DF2}, $\Mhalo = 10^{9.1-9.5}\Msun$ (with the range indicated by the error bar). Large samples of low-mass galaxies with stellar and halo mass estimates, measured effective radii, and GC counts will be essential for testing the predicted relation between these quantities in our model.

In simulated $L^*$ galaxies from the E-MOSAICS project \citep{Pfeffer18,emosaicsII} the number of GCs $N_{\rm GC}$ is found to correlate with the formation time of the DM halo, such that earlier-forming haloes host larger GC populations \citep[][Table B1]{emosaicsII}. This trend is naturally reproduced in our model via the assumption that the time-averaged SFR at $z\ga2$ was higher in haloes that collapsed earlier. For a fixed galaxy size $\re$, the increase in $\NGC$ directly follows from the observed increase in both the cluster formation efficiency (i.e. the fraction of stellar mass in bound clusters), and the ICMF truncation mass $\Mmax$, with $\SigmaSFR$ \citep{Adamo20a}.

To quantify the abundance of GCs relative to field stars, Figure~\ref{fig:global_GCs2} shows the GC specific frequency (Eq.~\ref{eq:T_N}) for GCs with $M>10^4\Msun$. The general trend of increasing specific frequency with decreasing stellar mass (for lines parallel to the cosmic baryon fraction) predicted by the model matches qualitatively the dominant observed trend in the \citet{Forbes18b} sample \citep[see also][]{Peng08,Georgiev10}. The model overpredicts the present-day observed $\TN$ in the most massive haloes, but this is not a problem since observations only account for the surviving clusters. In fact, \citet{Kruijssen15b} predicts that disruption is a key driver of the observed relation between $\TN$ and galaxy mass, and can suppress it by a factor of $10-100$ in $L_*$ galaxies \citep[see also][]{Bastian20}. The specific frequency in the model is relatively constant for galaxies on the mean SMHM relation, and increases steeply (and then decreases) with stellar mass at fixed $\Mhalo$, reaching the highest values at $\Mhalo = 10^7-10^8\Msun$. This occurs because the high gas pressure environments of galaxies with larger $\Mstar$ at fixed $\Mhalo$ generally produce a larger fraction of their stars in massive bound clusters compared to galaxies near the SMHM relation. This trend may also explain the increased specific frequencies of low-mass galaxies in the central region of the Virgo cluster, which \citet{Peng08} argue is a predictable result of their relatively early formation times compared to galaxies in the outskirts. Our model also predicts a drop in the specific frequency $\TN$ for galaxies near the cosmic baryon fraction due to the rapid increase in the minimum cluster mass (see Figure~\ref{fig:global_GCs1}). However, mass loss in clusters with $M<10^5\Msun$ may modify this trend.

\subsection{Galactic structure}
\label{sec:structure}

\begin{figure*}
    \includegraphics[width=\textwidth]{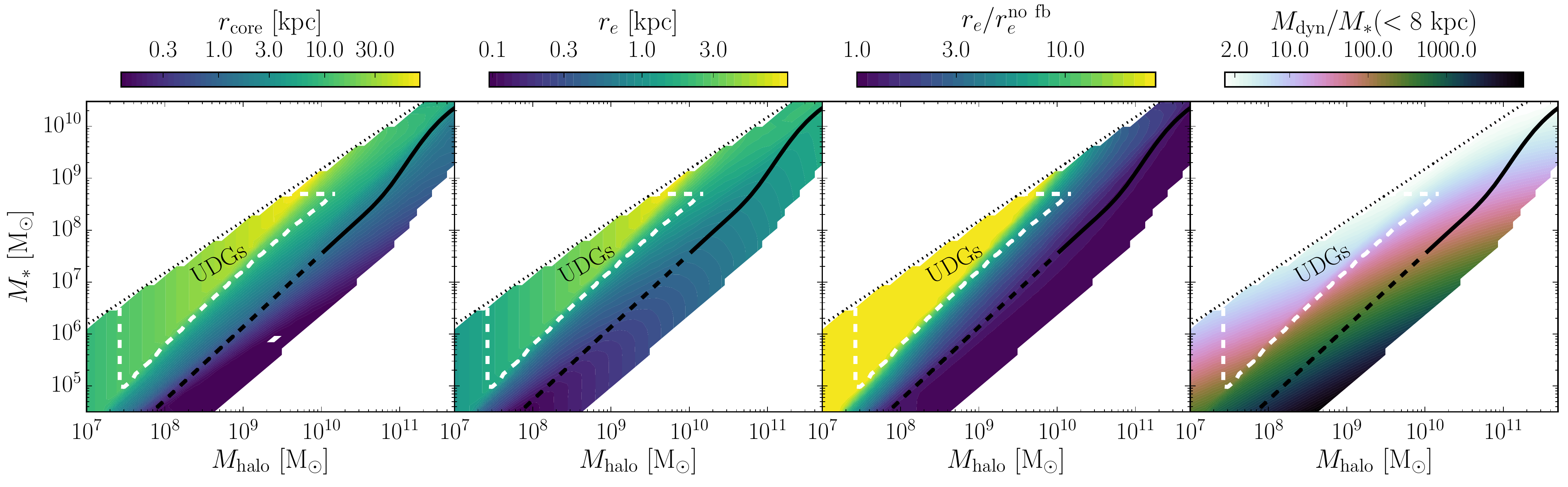}
    \caption{Impact of scatter in the SMHM relation on galaxy and DM halo structural properties at $z=0$. From left to right the panels show the size of the feedback-induced DM profile core, the galaxy effective radius, the increase in the effective radius relative to the case where there is no feedback-induced halo expansion, and the dynamical mass-to-stellar mass ratio within a radius of $8\kpc$, respectively. The region defined by the white dashed line corresponds to the commonly adopted observational definition of UDGs, $\re>1.5\kpc$ for $\Mstar\la5\times10^8\Msun$. The solid line is the mean SMHM relation from \citet{Behroozi13}, and the dotted line indicates the cosmic baryon fraction. Galaxies with the largest upward scatter in $\Mstar$ at fixed $\Mhalo$ can form DM cores with $\rcore \gg 10\kpc$, and expand their stellar components dramatically, reaching effective radii up to $\re\sim10\kpc$. The model predicts that UDGs \emph{with a broad range of dynamical mass-to-light ratios} form when the upwards scatter in $\Mstar$ at fixed $\Mhalo$ is $\gtrsim 1$ dex.}
\label{fig:global_structure}
\end{figure*}

The predicted structural properties of galaxies at $z=0$ are shown in Figure~\ref{fig:global_structure}. From left to right, the panels show the size of the DM core, the galaxy effective radius, and the increase in the effective radius relative to the case with no feedback-driven expansion. Galaxies on or below the mean SMHM relation do not expand significantly relative to the case where SN feedback is not included, and reproduce the observed relation between halo mass and galaxy size from \citet{Kravtsov13}. On the contrary, galaxies located more than $\sim 0.5$ dex above the relation inject a larger fraction of their SN energy into galactic winds, increasing their effective radius by a factor of $\gtrsim 10$ as the stellar mass increases at fixed $\Mhalo$. The transition into the expansion regime occurs closer to the mean SMHM relation in lower mass dwarfs. Galaxies in which SN feedback is most efficient at expanding the DM halo and stars, are generally those with the largest GC abundance per unit stellar mass, $\TN$.

The region inhabited by UDGs according to the common observational definition, $\re>1.5\kpc$ for $\Mstar\la5\times10^8\Msun$, is indicated by the white dashed line in Figure~\ref{fig:global_structure}. The model predicts that isolated UDGs will form from galaxies hosted by haloes with $\Mhalo \ga 10^{7.5}\Msun$ that scatter upwards from the SMHM relation by $\ga 0.5$ dex. Together with the elevated specific frequencies in the same region of the SMHM plane  (see Figure~\ref{fig:global_GCs2}), this provides a natural explanation for the higher GC abundance and specific frequency observed in low surface brightness galaxies compared to normal dwarfs at fixed stellar mass \citep{vanDokkum17,Lim18,Prole19a,Lim20}.

\subsection{The role of the degree of clustering of star formation in galaxy evolution}
\label{sec:clustering_role}

\begin{figure*}
    \includegraphics[width=\textwidth]{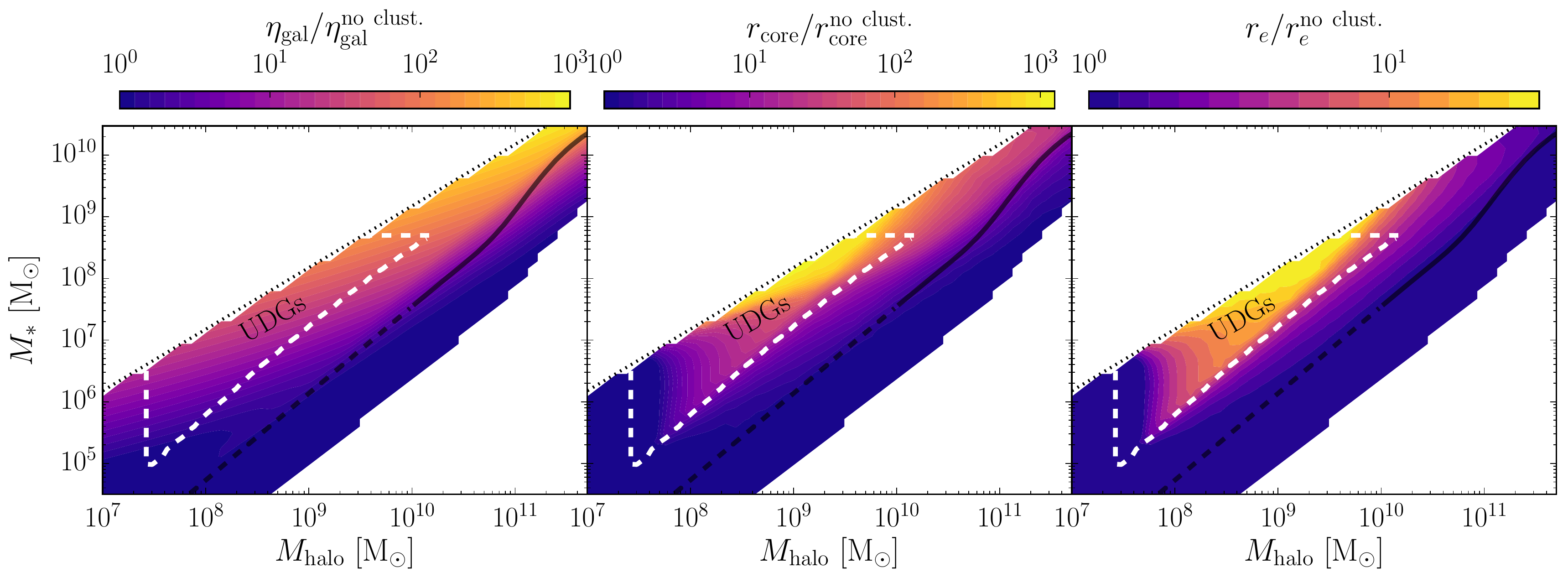}
    \caption{Impact of stellar clustering at the peak GC formation epoch on DM halo and galaxy structural evolution. The panels show the wind energy loading (left), the DM core size (middle), and the galaxy effective radius (right) at $z=0$ for the fiducial model relative to a model where clustering of SNe does not influence the driving of galactic winds (i.e.~$\fcl=1$). The solid line is the mean SMHM relation from \citet{Behroozi13}, and the dotted line indicates the cosmic baryon fraction. The enhanced coupling of SN energy to galactic winds due to overlapping SNe in star clusters is the dominant effect driving the expansion of the DM haloes and stellar discs of low-mass galaxies that scatter above the SMHM relation. As a result, our model predicts that feedback from massive star clusters is a necessary ingredient for explaining the origin of isolated ultra-diffuse galaxies.} 
\label{fig:clustering_effect}
\end{figure*}

Scatter above the mean SMHM relation influences the structure of model galaxies in two main ways. First, galaxies that scatter above the relation have larger stellar masses at fixed halo mass, and therefore more total SN energy available for winds to expand the DM and stellar components. Second, these galaxies formed in earlier collapsing haloes with larger SFR surface densities at $z=\zcoll$ due to higher gas pressures. This results in an increase in both their bound fraction and mean cluster mass, which in turn increase the spatial and temporal clustering of SNe and the energy loading of galactic winds. To understand which of these two effects is the dominant driver of the expansion of the galaxy and halo, we show in Figure~\ref{fig:clustering_effect} the wind energy loading $\etagal$, the DM core size $\rcore$, and the galaxy effective radius $\re$, of the fiducial model compared to a model with no SN clustering (setting $\fcl=1$ in Equation~\ref{eq:eta_E}). 

It is evident from Figure~\ref{fig:clustering_effect} that the spatial and temporal clustering of SNe in massive clusters is the dominant mechanism that determines galactic wind energies and galaxy structure in those galaxies above the SMHM relation which exhibit the largest amounts of feedback-induced expansion, $\Mhalo \approx 10^{7.5}-10^{10}\Msun$. Including the degree of clustering of stellar feedback sources in the model has a large impact on galactic outflows: for galaxies that scatter $\ga 0.5$~dex above the SMHM relation it increases wind energies by a factor of up to $\approx 1000$. For these galaxies, the middle and right panels of  Figure~\ref{fig:clustering_effect} show that the enhanced energy loading of galactic outflows due to clustered feedback strongly influences galaxy evolution: it increases the DM core sizes by up to a factor of $\sim 1000$, and galaxy effective radii by up to a factor of $\sim 30$ compared to the case with no SN clustering. This implies that a treatment of the clustering of feedback sources could be essential for future galaxy formation models aiming to reproduce the full diversity of the low-mass galaxy population.

\section{Implications for the formation of UDGs}
\label{sec:UDGs}

A large fraction of the region above the mean SMHM relation appears to fall within the regime of observed UDGs (see Fig.~\ref{fig:global_structure}). The model predicts that isolated UDGs form from dwarf galaxies that scatter $\ga 1$ dex above the mean SMHM relation as a result of the relatively early assembly of their host DM haloes, and its effect on stellar clustering and feedback-driven expansion (see Section~\ref{sec:clustering_role}). The last panel of Figure~\ref{fig:global_structure} shows that UDGs in the model have a broad range of dynamical mass-to-stellar mass ratios, $\Mdyn/\Mstar \approx 1-100$, when measured well outside the extent of their stellar component at $r=8\kpc$. 

In this section we compare the predictions of the model to the detailed properties of some of the most extreme outliers from the SMHM relation in Figure~\ref{fig:SMHM}, NGC1052-DF2 and DF4, and six isolated gas-rich UDGs with extremely high baryon fractions $\fbar \equiv \Mbar/\Mdyn \sim 1$. Because their mass distribution is known at very large galactocentric radii ($r \sim \R200$), their halo masses can be measured directly. The low DM content of these objects (consistent with zero in some cases) therefore presents the greatest challenge to galaxy formation in the \LCDM~ context, in which DM is always dominant at scales $r\ga\R200$.

\subsection{The case of NGC1052-DF2 and DF4}
\label{sec:DF2}

\begin{figure*}
    \includegraphics[width=0.9\textwidth]{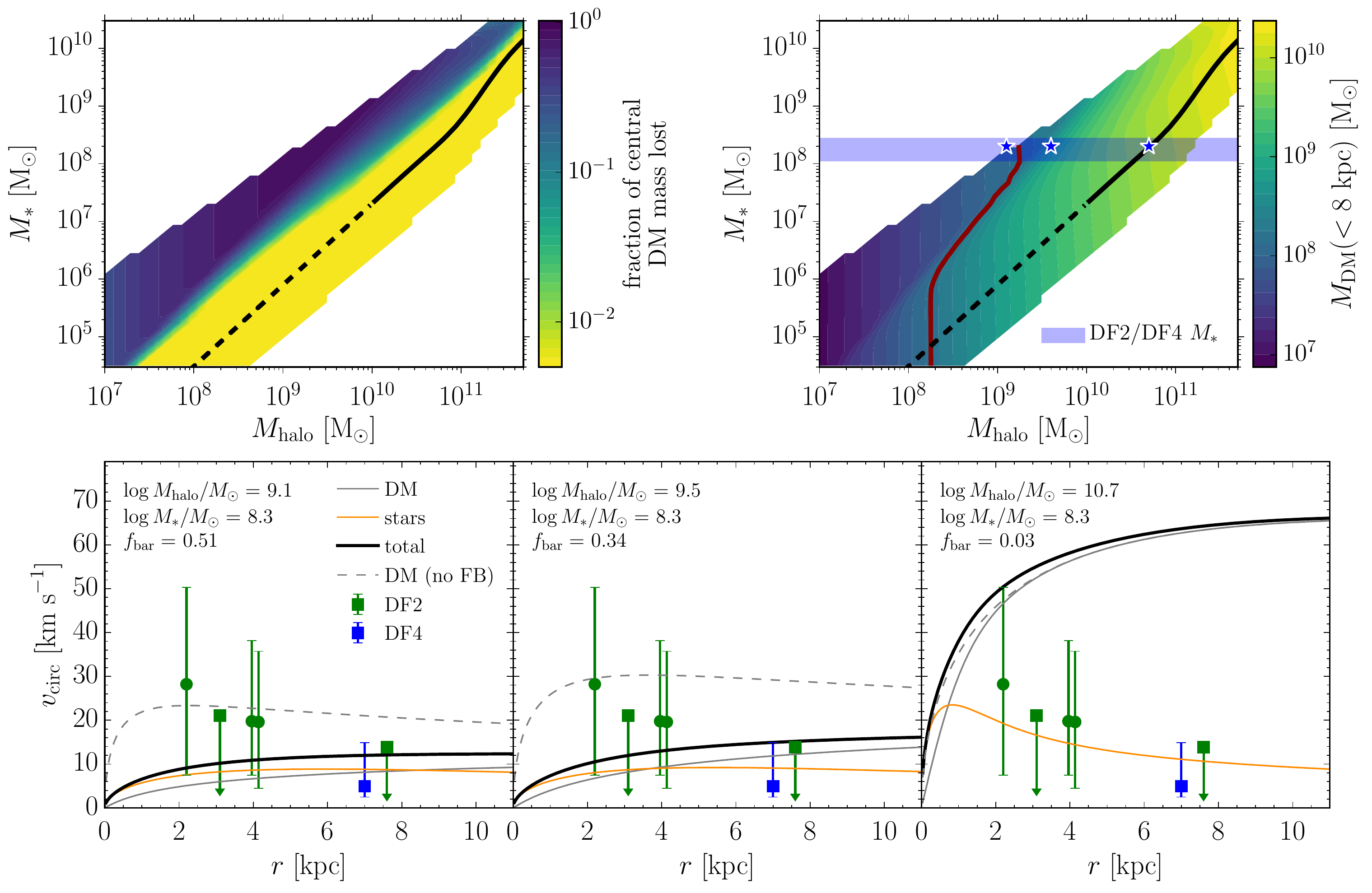}
    \caption{Impact of feedback-driven expansion on the density profile of galaxies as a function of their position in the $\Mstar-\Mhalo$ plane. First row: amount of DM mass lost from the inner $8\kpc$ relative to the case with no feedback (left), and DM mass within $8\kpc$ (right). The range of stellar mass for DF2 and DF4 including uncertainties is indicated by the horizontal shaded band, and the observational upper limit on the enclosed DM mass, $\MDM(<8\kpc)<1.5\times10^8\Msun$ \citep{vanDokkum18a}, is shown as a red line. The star symbols indicate the three halo masses considered in the bottom row. The solid line is the mean SMHM relation from \citet{Behroozi13}. Second row: predicted circular velocity curves for a galaxy with the stellar mass of DF2/DF4 ($\Mstar \approx1-2\times10^8\Msun$) and three different possible values of halo mass (from left to right) corresponding to the star symbols in the upper right panel. Points show the dynamical constraints for DF2 (green symbols) and DF4 (blue square). The halo and stellar masses, and the baryon fraction within $8\kpc$ are indicated in each panel. While the model on the mean SMHM relation (bottom right) is DM-dominated and well above the constraints, models with halo masses up to $\Mhalo \sim 3\times10^{9}\Msun$ satisfy both the DM mass constraints and the circular velocity curve at all radii.} 
\label{fig:DF2}
\end{figure*}

The predictions for a galaxy with the stellar mass of DF2/DF4, $\Mstar \approx 2\times10^8\Msun$, are presented in Figure~\ref{fig:DF2}. The top row shows how the formation of the core impacts the DM density profiles of galaxies when measured at a galactocentric radius $r=8\kpc$, which approximately corresponds to the outermost dynamical constraints for DF2 and DF4. The left panel shows the DM mass deficit, i.e.~the fraction of DM mass that was removed by feedback-driven expansion within $8\kpc$. Interestingly, the DM mass deficit is below 1 per cent for galaxies near or below the mean SMHM relation, and increases steeply to $\gtrsim 10$ per cent above it. Galaxies with masses $\Mhalo \sim 10^9 - 10^{10}\Msun$ that lie more than $\sim 1 {\rm dex}$ above the mean SMHM relation reach the largest DM deficiencies, nearly 100 per cent, due to their large massive cluster populations (see Figure~\ref{fig:global_GCs1}), and the resulting high clustering of SN events (see Figure~\ref{fig:model4}). The right panel shows the DM mass enclosed within $8\kpc$, and compares it to the possible locations of DF2 and DF4 on the $\Mstar-\Mhalo$ plane, given that their total halo mass is not known (shaded band). To provide an upper limit on $\Mhalo$, we assume no DM has been stripped. The red line indicates the upper limit on the enclosed DM mass from \citet{vanDokkum18a}. The small region left of the red line that overlaps with the shaded band represents the part of the SMHM plane where DF2 and DF4 could have formed. It corresponds to a DM halo with $\Mhalo \la 2\times10^9\Msun$ which contains nearly its entire cosmic baryon fraction in stars, and lost $\ga 90$ per cent of its central DM mass due to feedback-driven expansion. 
To explore the effects of the variation in the $\Mstar/\Mhalo$ ratio, the bottom row of Figure~\ref{fig:DF2} shows the circular velocity profiles corresponding to the positions of the three star symbols in the upper right panel. The halo and stellar masses, and the enclosed baryon fraction within $8\kpc$ are indicated in each panel. Dynamical constraints for DF2 \citep{vanDokkum18a,Emsellem19}, and DF4 \citep{vanDokkum19} are shown for comparison. Evidently, the model at the mean SMHM relation (right panel) fails to reproduce the dynamical constraints due to its massive DM halo with a relatively small core. The two models that scatter by more than 1 dex above the SMHM relation (left and middle panels) satisfy the dynamical constraints due to a combination of two factors: a lower mass host DM halo with a lower circular velocity $\vcirc$, and stronger winds due to increased stellar clustering, which more easily carve a very large DM core in a halo of lower central density and binding energy. The models that reproduce the mass profiles predict effective radii $\re \sim 5\kpc$, larger than the observed values for DF2 and DF4, $\sim 2\kpc$. Due to the many simplifying assumptions in the model, we do not expect this to be a major shortcoming of the model. 

\subsection{Isolated gas-rich UDGs}
\label{sec:field}

As Figure~\ref{fig:SMHM} clearly shows, NGC1052-DF2 and DF4 are not unique in their apparent DM deficiency. Many galaxies from several samples including the nearby dwarfs from \citep{Forbes18b}, the SPARC database \citep{Lelli16,Li20}, and many other isolated low-mass galaxies in the Local Volume and beyond  \citep[e.g.][]{geha06,Ferrero12,Oman16,Klypin15,Papastergis15a,Schneider17,Trujillo-Gomez18} seem to inhabit very low mass DM haloes relative to the expectation from the mean SMHM relation. Similar examples of this phenomenon are also found among massive early type galaxies \citep{Padmanabhan04}, and massive disc galaxies \citep{Posti19}. 

When the dynamical masses of DF2 and DF4 were first estimated, they appeared to be the most extreme examples of DM-deficient galaxies, with $\fbar \sim 1$ within a radius of $\sim 8\kpc$ (see Fig.~\ref{fig:SMHM}). However, in a recent study, \citet{ManceraPina19,ManceraPina20} found six gas-rich isolated UDGs for which the rotation velocity of the \HI~ gas is very slow compared to non-UDGs with the same baryonic mass. As a result, the stellar and cold gas content of these galaxies accounts for most (and sometimes all) of their dynamical mass within the extent of their \HI~ disc, $\sim 8-11\kpc$. Through a careful analysis of the uncertainties, \citet{ManceraPina20} rule out systematics in the derivation of the rotation curves as a source of this effect. We build simple mass models of these UDGs and show the inferred halo masses in Figure~\ref{fig:SMHM} \footnote{See Appendix~\ref{sec:massmodels} for a description of the mass models.}, highlighting the striking similarity with DF2 and DF4 in terms of their DM deficiency. Four out of the six isolated UDGs in this sample have baryon fractions above the cosmic value, and three of these (indicated by upper limits on their halo masses) do not require any DM to explain their dynamics at large radii.

\begin{figure*}
    \includegraphics[width=0.85\textwidth]{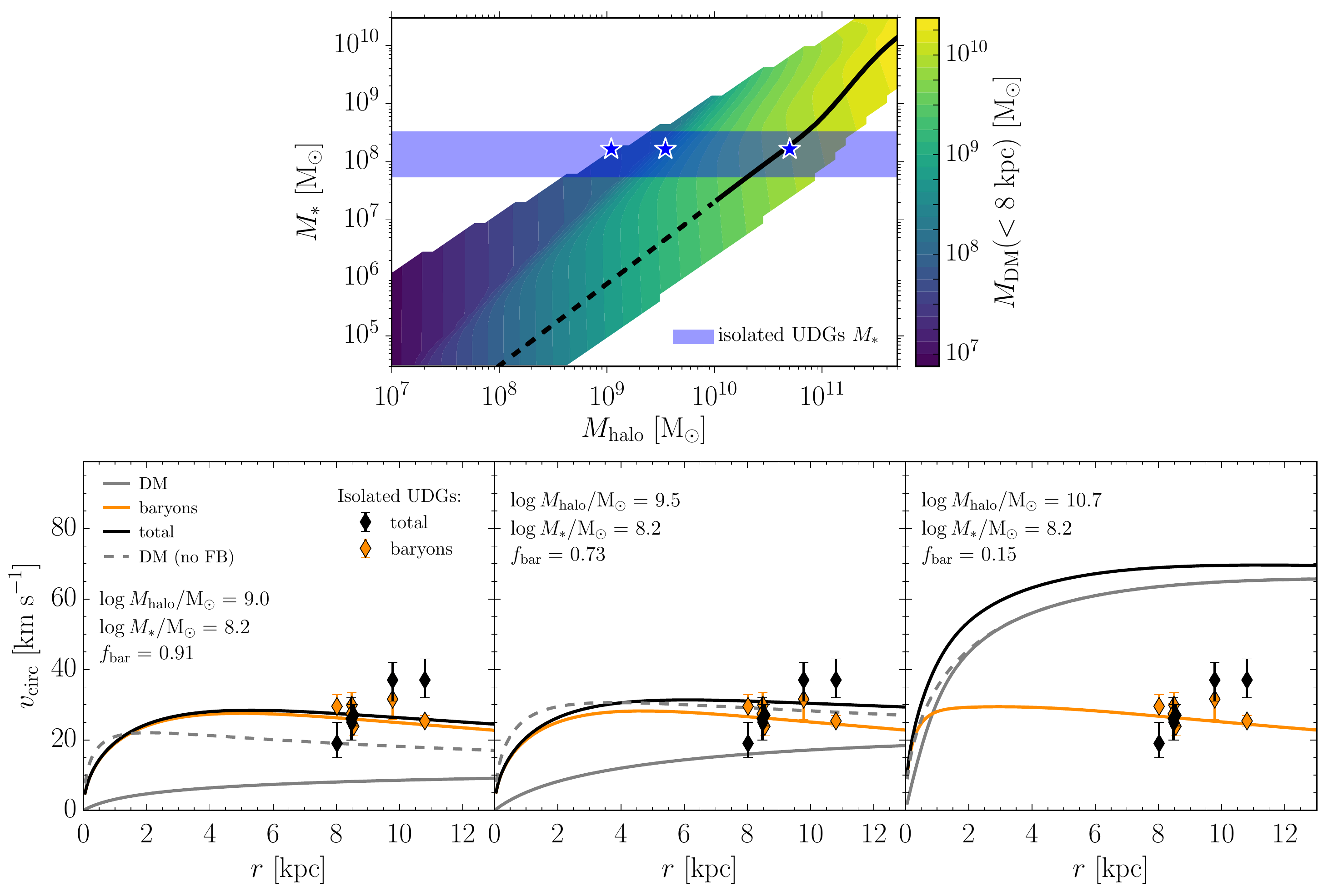}
    \caption{Predicted DM content and mass profiles at large radii compared to observations of gas-rich isolated UDGs. Top: enclosed DM mass within the central $8\kpc$ as a function of position in the $\Mstar-\Mhalo$ plane. The blue shaded band indicates the range of stellar masses of the isolated UDG sample from \citet{ManceraPina20}. The star symbols indicate the location of the three galaxy models shown in the bottom row. The solid line is the mean SMHM relation from \citet{Behroozi13}. Bottom left and middle: mass profiles of two model galaxies that fit the gas-rich UDG mass profile constraints. Bottom right: mass profile of a galaxy with the mean observed stellar mass of the UDG sample and the mean halo mass from the SMHM relation. Each of the bottom panels lists the corresponding halo mass, stellar mass, and baryonic fraction within $8\kpc$ $\fbar$, and shows the baryonic and total observed circular velocities from \citet{ManceraPina20} as points with error bars. The $\sim 2$ dex upwards departure from the mean SMHM relation (bottom left panel) produces more efficient feedback-driven expansion of the DM and stars, resulting in dynamical dominance of the baryonic mass within the inner $8-11\kpc$. A galaxy with the mean halo mass (bottom right panel) fits the baryonic rotation velocity but overpredicts the total $\vcirc$. Since the model does not predict gas properties, we assumed exponential gas density profiles with a scale-length set by the mean stellar scale-length of the UDG sample.} 
\label{fig:field_UDGs}
\end{figure*}

By construction, our model accounts for galaxies with baryon fractions as high as the cosmic mean. However, objects that lie above it and have extremely low DM content can potentially be explained by the effect of feedback-driven DM expansion due to enhanced stellar clustering in galaxies above the mean SMHM. As in the case of DF2 and DF4, this could increase their baryon fractions to $\fbar \equiv \Mbar/\Mdyn \sim 1$.  Figure~\ref{fig:field_UDGs} shows the predictions for galaxies with properties in the range spanned by the isolated UDGs from \citet{ManceraPina20}, with baryonic masses $1.1\times10^9\Msun< \Mbar <2.3\times10^9\Msun$ and baryon fractions $0.47< \fbar(\la 8\kpc) <1$. The top panel shows the DM mass of galaxies enclosed within $r=8\kpc$ in the $\Mstar-\Mhalo$ parameter space. Since the model does not predict the cold gas properties at $z=0$, we simply assume an exponential gas density profile with a scale-length equal to the mean disc scale-length of the sample in order to facilitate comparison with our predictions. The blue shading indicates the range of stellar mass of the observed UDGs. The bottom left and middle panels show the predicted circular velocity profile for an object in this region with $\Mhalo = 1-3\times10^{9}\Msun$ and the mean stellar mass of the isolated UDG sample, $\Mstar = 1.6\times10^8\Msun$. Although these model galaxies should contain significant amounts of DM in the central $\sim 10\kpc$ (the `no feedback' case shown with a dashed line), the increased clustering of star formation and wind driving efficiency have carved a very large core $\rcore > 25\kpc$ in their DM profiles (solid grey line). The corresponding reduction in the inner DM density, together with the increased baryon mass due to the early collapse of the halo, lead to the large enclosed baryon fraction $\fbar > 0.73$ within $8\kpc$. Also note that the model predicts very large disc effective radii $\re = 3.2-5.2\kpc$, in agreement with the observed range of the UDGs, $\sim 3-7\kpc$. For comparison, the bottom right panel shows the profile of a galaxy with the same stellar mass but with a halo mass determined by the mean SMHM relation, $\Mhalo = 5\times10^{10}\Msun$. In this case the baryonic circular velocity profile fits the observed galaxies, but the total circular velocity is significantly overpredicted, and the central baryon fraction is only $0.15$ due to the large halo mass and relative inefficacy of SN feedback. 

$\HI$-bearing UDGs are commonly found in the field. They represent about 6 per cent of all low-mass galaxies with $8.5< \log\MHI/\Msun <9.5$, and have a cosmic abundance similar to that of cluster and group UDGs \citep{Jones18}. \citet{Leisman17}, \citet{Jones18}, and \citet{Guo20} find that these galaxies generally rotate slower than normal galaxies of similar baryonic mass, indicating that DM-deficiency may be a general property of field UDGs. This raises the intriguing possibility that a single formation channel, like the one proposed here, could account for the structural properties of both field and cluster/group UDG populations. This mechanism may also remove the need for intense tidal stripping for explaining UDGs in dense environments.

\section{Inferred halo masses and baryon fractions}
\label{sec:obs_halomass}

\begin{figure*}
    \includegraphics[width=1.4\columnwidth]{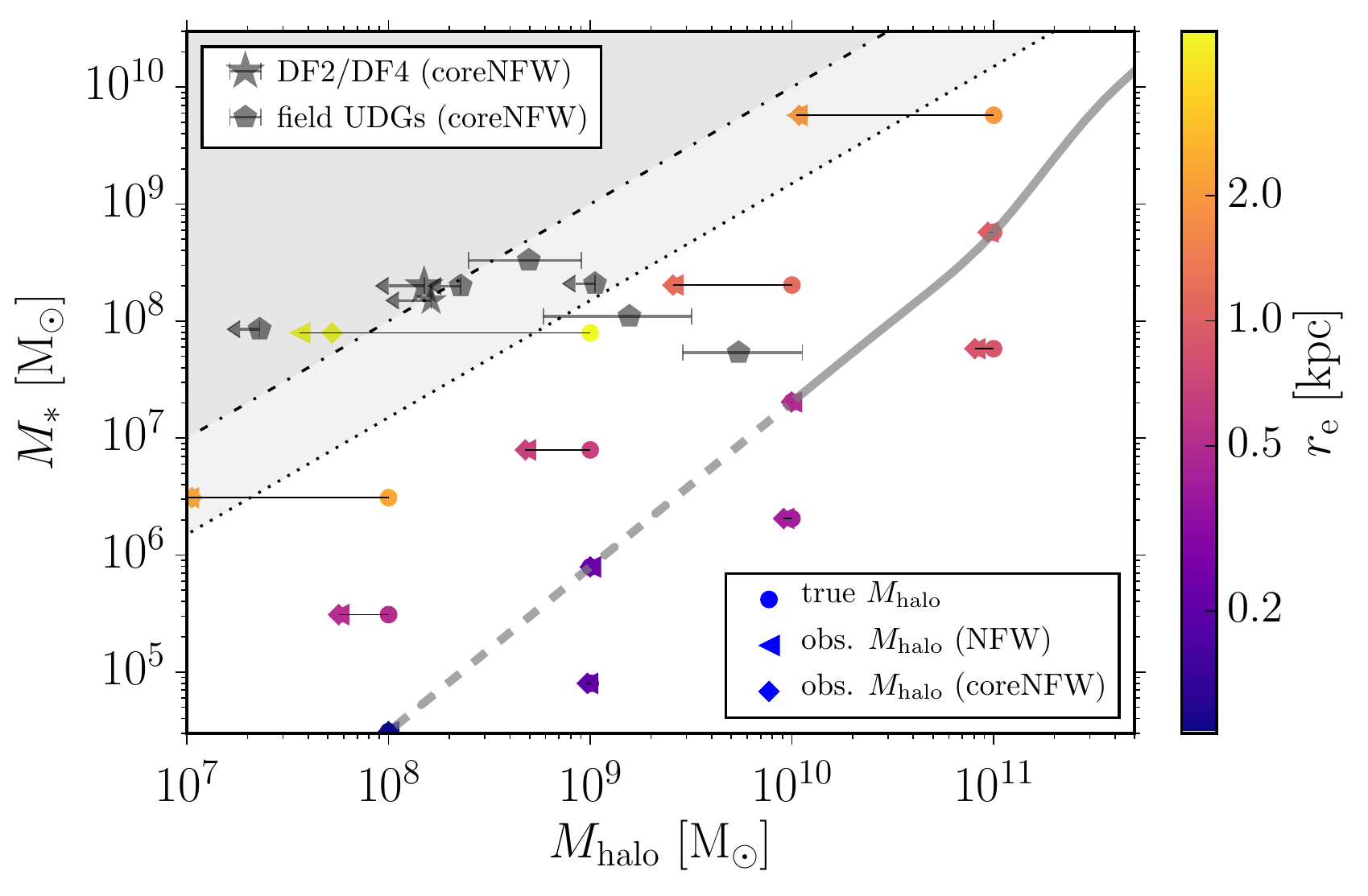}
    \caption{Impact of feedback-driven expansion on the inferred halo mass of observed galaxies. Circles show the true halo mass of model galaxies with varying amounts of scatter from the mean SMHM relation at various values of $\Mhalo$. Lines connect the true $\Mhalo$ to the value obtained by fitting an NFW (triangles) or a coreNFW (diamonds) profile to the DM circular velocity measured at a large radius, $r=8\kpc$. The dotted and dash-dotted lines indicate the cosmic baryon fraction and the $\Mstar=\Mhalo$ line, respectively. The symbols are coloured to indicate the galaxy effective radius. The grey symbols with error bars reproduce the dynamical constraints for DF2, DF4, and gas-rich isolated UDGs from Figure~\ref{fig:SMHM}. The solid line is the mean SMHM relation from \citet{Behroozi13}, the dotted line shows the cosmic baryon fraction, and the dash-dotted line indicates $\Mstar=\Mhalo$. Assuming an NFW profile to fit the enclosed dynamical mass of strongly expanded galaxies severely underestimates their true $\Mhalo$, and shifts their apparent position towards the left, making them appear very DM-deficient, with $\Mstar>\Mhalo$ in some cases. The low inferred halo masses of DF2/DF4 and some field UDGs could be explained by this effect.  } 
\label{fig:SMDM}
\end{figure*}

The modification of the mass distribution by stellar feedback for galaxies above the mean SMHM relation could have important consequences for dynamical models aimed at inferring the mass of the host DM halo from the stellar or gas kinematics. Figure~\ref{fig:SMDM} shows the effect of fitting two commonly assumed DM halo density profiles to observations that determine the circular velocity at a radius of $r=8\kpc$. This radius is representative of the most extended dwarf galaxy and field UDG gas rotation curves \citep{Lelli16,ManceraPina20}, as well as the outermost GC dynamical measurements in DF2 and DF4 \citep{vanDokkum18a,vanDokkum19}. 

The figure shows that galaxies which scatter upwards from the mean SMHM relation and become very expanded due to the enhanced clustering of SN feedback, have such large DM cores that naive fits to the kinematics at $r\la 8\kpc$ assuming the NFW profile result in a large underestimation of the true cosmological halo mass (i.e. the halo mass without expansion). For galaxies near the cosmic baryon fraction, this effect can result in an underestimation of the true halo mass by more than an order of magnitude. Interestingly, \citet{ManceraPina20} find that the sparsely-sampled rotation curves of their gas-rich UDGs (also reproduced in Figure~\ref{fig:SMDM}) can only be fit with an NFW profile of extremely low concentration compared to the mean expected in \LCDM~ haloes. When fit using NFW profiles (which do not account for stellar feedback), the effect of enhanced DM expansion in galaxies that scatter above the SMHM relation could explain both the \emph{apparent} position of these galaxies in the $\Mstar-\Mhalo$ plane, and their seemingly low concentrations. Figure~\ref{fig:SMDM} also suggests that the apparently large scatter in $\Mstar$ at fixed $\Mhalo$ in the dwarf galaxy samples in Figure~\ref{fig:SMHM} could be enhanced in large part by underestimation of the true halo mass. This would push their apparent positions above the cosmic baryon fraction, with $\Mstar>\Mhalo$ in the most extreme cases, and in contradiction with \LCDM~ expectations. We therefore expect that the real scatter in $\Mstar(\Mhalo)$ may be considerably smaller\footnote{By extension, we also expect the true scatter on the relation between the number of GCs and the DM halo mass to be smaller than observed \citep[see e.g.][]{Forbes18b,Bastian20}.}. Detailed modelling of observed rotation curves in large galaxy samples is needed to constrain the model and to provide an estimate the true scatter in $\Mstar/\Mhalo$.

\section{Discussion}
\label{sec:discussion} 

Our model predicts that low-mass galaxies that scatter significantly above the SMHM relation formed earlier, and experience more clustered star formation and more energetic outflows, causing them to expand enough to become UDGs. This intrinsic channel for UDG formation has also been suggested by numerical studies \citep{DiCintio17,Jiang19,Jackson20a}. A novel result of our analysis is that the growth of the DM core in UDGs can be so pronounced for objects with the highest $\Mstar/\Mhalo$ ratios (near the cosmic baryon fraction), that the DM mass enclosed within $\sim 10\kpc$ is extremely reduced. In addition, the model simultaneously explains how the early formation of these galaxies leads to an excess in the number of GCs. These predictions are strikingly similar to the puzzling observations of NGC1052-DF2 and DF4, and of six isolated gas-rich UDGs, which all seem to have very low DM content ($\la 50$ per cent of the total mass) at large radii, where DM is expected to dominate. The model therefore accounts for the formation of galaxies with low DM content in the field, as well as in groups and clusters, without the need for highly eccentric orbits. 

\citet{Trujillo-Gomez20b} used the GC mass function of the 11 unusually massive GCs in DF2 to reconstruct its galactic environment $\sim 9\Gyr$ ago. Due to the implications of the peculiar GC mass function for the size of the star-forming region, they conclude that the massive GCs likely formed during a major merger. Our model does not explicitly include the effect of mergers. However, we expect the overall expansion of the galaxy to be mostly insensitive to the details of how the GCs were produced (i.e.~by rapid gas accretion or by a major merger), such that the results of the merger model are complementary to the ones presented here.   

We find that the observed large scatter in the SMHM relation of low-mass galaxies can originate in part from the combined effects of early collapse and increased stellar clustering on the central mass distribution. When mass models based on DM-only simulations are fitted to the dynamics of galaxies with the largest upward scatter, the strong expansion of the DM results in an underestimation of the true (pre-expanded) halo mass. This could help explain the objects that lie above the cosmic baryon fraction, and especially those with $\Mstar>\Mhalo$ (see Figure~\ref{fig:SMDM}). The model also qualitatively reproduces the higher GC specific frequency in galaxies with lower surface brightness, found across dwarfs, UDGs, and low surface brightness galaxies in galaxy clusters \citep{Lim18,Prole19a,Lim20}. Because UDGs are more commonly found near the cluster core, and galaxies in this region formed earlier \citep{Peng08}, our model predicts that they should have scattered above the mean SMHM relation, causing them to form more GCs and expand. 
 The existence of DM-deficient isolated UDGs, as well as many other field galaxies with low DM content (i.e. large baryon fractions; see Section~\ref{sec:field}) indicates that \emph{the lack of DM is not exclusively caused by stripping in a group environment} as proposed by many numerical studies \citep[e.g.][]{Ogiya18,Carleton18,Nusser20,Maccio20,Jackson20b}. Instead, intrinsic mechanisms acting on isolated galaxies, such as enhanced outflows due to stellar clustering as proposed here, are needed to explain this phenomenon. Most field UDGs are blue and star-forming \citep{Prole19b}. Furthermore, gas-rich field UDGs represent a substantial fraction of the galaxy population, and tend to rotate slower than normal galaxies of the same baryonic mass \citep{Jones18}. This suggests that DM-deficiency might be generic to all field UDGs that scatter well above the SMHM relation, and highlights the need to understand these objects. Furthermore, the model predicts that galaxies with $\Mstar\sim 10^5-10^7\Msun$ could constitute a large fraction of the UDG population that is yet to be discovered. \citet{Silk19} proposed a mechanism for removing DM and forming massive GCs in high-speed collisions between dwarf galaxies in gas-rich proto-groups. These predictions were confirmed by \citet{Lee21} using hydrodynamical simulations, showing that the collision results in the formation of a gas-poor compact merger remnant and several massive GCs. While DF2 and DF4 could be consistent with forming in high speed collisions, the six gas-rich DM-deficient UDGs studied by \citet{ManceraPina20} cannot be explained by this mechanism due to their high gas content, ongoing star formation, and isolation. \citet{Shen21} recently estimated the relative distance between DF2 and DF4 to be $2.1\pm0.5\Mpc$. This large separation implies that at least one of the two galaxies is isolated, and likely rules out environmental formation mechanisms, including high-speed collisions.

While our model does not account for environmental effects, it predicts an large intrinsic diversity (at formation) in the sizes and mass-to-light ratios of field galaxies (Fig.~\ref{fig:global_structure}). This intrinsic diversity would only be enhanced by the effects of ram pressure stripping of gas and tidal stripping of stars and DM on the broad distribution of satellite infall times and trajectories. Detailed predictions for these effects should be made using numerical simulations with initial conditions provided by the model presented here. Another element that the model does not include is the evolution of the cold gas content. While we assume here that isolated galaxies retain their cold gas until $z=0$, it is possible that in certain cases a strong clustered feedback episode could remove all the gas permanently, effectively quenching the galaxy. Suggestive evidence of this process may be found in the first study of the spatially-resolved stellar populations in a UDG.  \citet{Villaume21} studied DF44, a UDG in the outskirts of the Coma cluster, and determined that the galaxy has a very narrow SFH and a flat radial stellar metallicity and age gradient. They conclude that this is consistent with DF44 having formed entirely, along with its large population of $\sim 100$ GCs\footnote{This estimate has been challenged by \citet{Saifollahi21} and \citet{Bogdan20}. \citet{Saifollahi21} estimate a much lower number, $\approx 20$ GCs, in DF44.} \citep{vanDokkum16_DF44}, in a single burst of star formation $\sim 10\Gyr$ ago. Since it is likely that DF44 was quenched before falling into Coma, this raises the possibility that the same SF burst that produced the large population of GCs and expanded the stars also quenched the galaxy $\sim 10\Gyr$ ago. In this scenario, DF44 formed near the mean SMHM relation \citep{vanDokkum19_DF44} and had an early strong burst of SF that expanded the stellar component to $\re=4.6\kpc$, flattening the age and metallicity gradients, and quenching the galaxy. Early self-quenching prevented subsequent generations of star formation (which are centrally concentrated due to the infall of fresh gas) from producing a young disc with a smaller effective radius (see Eq. \ref{eq:r_e_total}). The lack of a young stellar component could therefore preserve the flat gradients caused by expansion.  

Our results also show that due to their higher GC occupation, UDGs may be offset from the linear $\Mgc-\Mhalo$ relation observed in normal galaxies. This suggests that using this relation to infer the DM halo mass of UDGs may be seriously biased \citep[also see the discussion in][]{vanDokkum18b}. The high values of $\NGC$ in cluster UDGs would therefore place them in the more massive DM haloes of normal galaxies with the same GC occupation, creating the illusion of UDGs that scatter downwards from the mean SMHM relation. Interestingly,  this is the result that \citet{Prole19a} obtain for Fornax UDGs, assuming that they follow the $\Mgc-\Mhalo$ relation.

\section{Conclusions}
\label{sec:conclusions}

Through a simple semi-empirical model, we have explored the effect of significant scatter in the low-mass SMHM relation -- as suggested by dynamical probes -- on the star cluster populations, feedback-driven winds, and structural evolution of galaxies and their host DM haloes. To minimise the number of free parameters, the model uses empirical galaxy scaling relations, clustering constraints on the galaxy-halo connection, as well as the results of large hydrodynamic cosmological simulations, and detailed numerical models of the generation of galactic winds by clustered SNe. For simplicity, it accounts for mergers as part of the smooth matter accretion component. Due to the lack of constraints on some elements of the model, several assumptions must be made. For reasonable parameter choices, the results show that DM haloes that collapse earlier and therefore achieve larger SFRs at $z\ga2$, reach the high pressure ISM conditions necessary to produce a larger fraction of stars in bound clusters that are on average more massive. In general, the larger the departure of a galaxy upwards from the mean SMHM relation, the more clustered its star formation and therefore its SN feedback will be at the collapse epoch. This in turn increases the energy loading of galactic winds, and the energy injected into the DM orbits by potential fluctuations due to massive outflows. Assuming that a small constant fraction of the wind energy couples to the potential of the halo, we estimate that the DM core size increases steeply with the upwards scatter, with galaxies near the cosmic baryon fraction forming cores with sizes $\rcore \gtrsim 10\kpc$.

The stars that formed before the epoch of halo collapse expand along with the DM, and this has a strong effect on galaxy sizes and surface brightness at $z=0$. It allows galaxies with $\Mstar < 10^9\Msun$ that scatter significantly ($\gtrsim 0.5-1.0$ dex) above the SMHM relation to expand enough to become UDGs, and suggests that \emph{all} UDGs could in principle be formed through this mechanism.

Our conclusions are summarised as follows:
\begin{itemize}
    \item A simple semi-empirical model predicts that scatter around the mean SMHM relation has important consequences for the evolution of the structure and star cluster populations of galaxies and their host DM haloes.
    \item The model assumes that galaxies with larger than average $\Mstar/\Mhalo$ ratios are hosted by DM haloes that collapse earlier, allowing for larger SFR at early epochs (Figure~\ref{fig:model1}). Their higher SFR surface densities (Figure~\ref{fig:model2}) indicate higher ISM pressures and therefore increased stellar clustering and larger GC populations compared to galaxies on the mean relation (Figure~\ref{fig:model3}). The predicted trend of increasing GC numbers and specific frequencies with scatter from the SMHM relation qualitatively matches the observed trend in nearby dwarfs, and may explain why UDGs in clusters have more GCs than normal galaxies of the same stellar mass (Figs.~\ref{fig:global_GCs1} and \ref{fig:global_GCs2}).
    \item The degree of clustering of star formation in space and time has a strong impact on galaxy evolution. For galaxies that scatter upwards from the SMHM relation by $\ga 0.5-1$ dex, a steep increase in stellar clustering and in the mean cluster mass drives a significant increase in the wind energy loading due to clustered SNe. This boost increases to a factor of $> 100$ for galaxies with the largest $\Mstar/\Mhalo$ (Figure~\ref{fig:clustering_effect}). The higher SFR at the collapse epoch in these galaxies also increases the total SN energy available to power galactic winds. This suggests that the spatial and temporal clustering of star formation and feedback should be included as an essential ingredient in galaxy formation models.
    \item Outflows in galaxies above the SMHM relation are more energetic and can potentially expand the orbits of the DM and old stars by a factor of $>10$ at $z=0$ compared to those on the mean relation. UDGs are formed naturally in progenitors with $\Mhalo\ga10^{7.5}\Msun$ which scatter upwards by $\ga 1$ dex (Figure~\ref{fig:global_structure}). The model predicts that a large fraction of low surface brightness field galaxies with $\Mstar\sim10^5-10^7\Msun$ could still lurk in the dark, comprising the low-mass end of the UDG population that is yet to be discovered.
    \item Enhanced feedback due to SN clustering drives the formation of a large core in the DM density profile, with the maximum effect at $\Mhalo\sim 10^9-10^{10}\Msun$ and $\Mstar\sim 10^8\Msun$ (Figure~\ref{fig:global_structure}). Cores with sizes $\rcore > 10\kpc$ significantly reduce the DM content, and boost the baryon fraction within $\sim 10\kpc$. Galaxies with the largest upwards scatter at $\Mhalo=10^9-10^{10}\Msun$ fit the dynamical constraints of DM-deficient galaxies like DF2/DF4 and several isolated UDGs, providing a natural pathway for the formation of isolated galaxies lacking DM. This mechanism accounts for the formation of DM-deficient galaxies in the field, and in groups/clusters, without the need for highly eccentric orbits to remove the DM within $\sim 10\kpc$ (Figs.~\ref{fig:DF2} and \ref{fig:field_UDGs}). Feedback from a large population of GCs formed in a single early burst of star formation could potentially account for the old ages and flat metallicity and age gradients observed in DF44 (Section~\ref{sec:discussion}). 
    \item The model naturally explains the large diversity in the dynamical mass-to-light ratios measured well beyond the extent of the stellar disc of observed UDGs, with a predicted range of $\Mdyn/\Mstar(<8\kpc) \approx 1-100$ (Fig.~\ref{fig:global_structure}). This diversity is driven by the increase in the stellar mass and its effect on stellar (and SN) clustering and the expansion of the inner DM halo with growing scatter above the mean SMHM relation. While environmental effects such as DM stripping are not needed to reproduce the properties of UDGs in the model, they are expected to increase the diversity of UDGs in group and cluster environments.
    \item Fitting the mass profile of galaxies with large DM cores with commonly used density profiles (such as the NFW profile) results in a significant underestimation of the halo mass by more than an order of magnitude in objects with the largest $\Mstar/\Mhalo$. This provides an explanation for galaxies that appear to exceed the cosmic baryon fraction, and in the most extreme cases where $\Mstar \ga \Mhalo$ (Fig.~\ref{fig:SMDM}). It also means that the intrinsic  scatter around the SMHM relation is smaller than observed.
\end{itemize}

These predictions can be compared to recent cosmological simulations of galaxy formation which resolve dwarf galaxies in representative volumes. Recently, \citet{Jackson20a} used the NewHorizon simulation to identify low surface brightness galaxies and find that these form preferentially in high density environments which lead to earlier formation times and stronger feedback-driven expansion. The dominant uncertainty in our model lies in the parameters $\fstar$, $\fDM$, and $c(\Mstar,\Mhalo)$, which control the fraction of stars formed during the halo collapse epoch, the fraction of galactic wind energy that couples to the DM, and the relation between stellar mass and halo concentration, respectively (see Appendix~\ref{sec:parameters}). These parameters can be constrained by high-resolution cosmological simulations that model cluster formation and disruption, and include its effect on the driving of galactic winds. Including the effects of stellar clustering on feedback and galaxy structure will be key to understanding how galaxies form and evolve in the \LCDM~ paradigm.

%%%%%%%%%%%%%%%%%%%%%%%%%%%%%%%%%%%%%%%%%%%%%%%%

\section*{Acknowledgements}

The authors are grateful to the anonymous referee for a constructive review, and to Alexa Villaume, Aaron Romanowsky, and Ben Keller for insightful discussions and feedback. STG, JMDK, and MRC gratefully acknowledge funding from the European Research Council (ERC) under the European Union's Horizon 2020 research and innovation programme via the ERC Starting Grant MUSTANG (grant agreement number 714907). JMDK gratefully acknowledges funding from the Deutsche Forschungsgemeinschaft (DFG, German Research Foundation) through an Emmy Noether Research Group (grant number KR4801/1-1) and the DFG Sachbeihilfe (grant number KR4801/2-1). MRC gratefully acknowledges the Canadian Institute for Theoretical Astrophysics (CITA) National Fellowship for partial support. This work made use of the software packages {\sc numpy} \citep{numpy}, {\sc scipy} \citep{scipy}, {\sc matplotlib} \citep{matplotlib} and {\sc colossus} \citep{colossus}.

\section*{Data availability}

The data underlying this article are available within the article and in the references to published sources.

%\newpage

%%%%%%%%%%%%%%%%%%%%%%%%%%%%%%%%%%%%%%%%%%%%%%%%

\bibliographystyle{mnras}
\bibliography{merged}

%\begin{thebibliography}{99}
%\end{thebibliography}

\appendix

\section{Observational data and mass models}
\label{sec:massmodels}

This section describes the observational data and corresponding mass models for the galaxies shown in Figure~\ref{fig:SMHM}. \citet{Forbes18b} provide mass models of nearby Local Group and isolated dwarf galaxies. These include analysis of high quality \HI~ and H$_{\alpha}$ rotation curves of galaxies without GCs from \citet{Read16b}, and a compilation of published rotation curves for galaxies with GCs from other sources in addition to velocity dispersion measurements at $\re$ for a few dSphs. For some dIrr galaxies without published rotation curves, the halo mass is obtained from a fit to the maximum circular velocity $\vmax$ as derived from the \HI~ linewidth, which has been shown to be a good approximation \citep{Trujillo-Gomez18}. GC occupation data for the isolated galaxies was obtained from \citet{Georgiev10}, while for the Local Group dwarfs it was obtained from a variety of sources \citep[see][for details]{Forbes18b}. 

\citet{Li20} fitted mass models to the SPARC ({\it Spitzer} Photometry and Accurate Rotation Curves) database \citep{Lelli16} of high quality rotation curves and $3.6\mu{\rm m}$ {\it Spitzer} data for 175 late-type galaxies ranging in mass from dwarfs to giant spirals. The models used here assume the coreNFW profile \citep{Read16a} and uniform priors on the halo concentration in the range $0<c<1000$. To avoid systematics due to large inclination corrections, we follow \citet{Lelli16} and limit the sample to inclinations $i>30\deg$, and remove objects with poor quality rotation curves (i.e. flag $Q=3$). The stellar mass is calculated from the $3.6\mu{\rm m}$ luminosity assuming a mass-to-light ratio $\Upsilon_*=0.5$. This introduces a small but negligible inconsistency with the mass models, which fit the stellar mass distribution using log-normal priors for the mass-to-light ratio of the disc and bulge components centred at $\Upsilon_{\rm disc}=0.5$ and $\Upsilon_{\rm bulge}=0.7$, with a standard deviation of $0.1$~dex.

\citet{ManceraPina20} analysed low-resolution $\HI$ rotation curves of 6 gas-rich isolated UDGs found in the ALFALFA survey. They modelled the full 3D data cubes (accounting for the effect of beam-smearing) to obtain the inclination, and provide measurements of stellar and $\HI$ mass, effective radius, and circular velocity $\vout$ at the outermost point of the rotation curve, $\rout \sim 8-10\kpc$. The circular velocity of the DM component is obtained by removing the contribution of stars and gas, $\vDM^2 = \vcirc^2 - \vbar^2$, where $\vbar^2 = G(\Mstar + 1.33\MHI)/\rout$, and with the factor 1.33 accounting for the contribution from Helium. Since the rotation curves are too coarse to properly fit mass models, we performed a simple fit to only the outermost rotation velocity using the coreNFW density profile from \citet{Read16a} which includes the effect of feedback-induced DM cores. Since the density profile depends on halo mass and concentration, an additional constraint is needed. For this, we assume the cosmological concentration-mass relation and its 1$\sigma$ scatter from \citet{DuttonMaccio14} to obtain an estimate of the uncertainty in $\Mhalo$. For three of the objects, the baryonic mass exceeds the dynamical mass at $\rout$, and the halo mass is consistent with zero. To obtain the upper limits on $\Mhalo$ shown in Figure~\ref{fig:SMHM} for these objects, we use the minimum integer multiple of the 1-$\sigma$ uncertainty in $\vcirc$ which results in a non-zero halo mass. 

\citet{Guo20} searched the ALFALFA \citep{Haynes11} catalogue for galaxies with luminosities $M_r > -18$ where the baryonic mass accounts for more than 50 per cent of the dynamical mass within the $\HI$ radius. For the 19 objects they find, they fit NFW mass models to the inclination-corrected rotation velocity obtained from the 20 per cent velocity width of the $\HI$ line profile. To avoid underestimation of $\Mhalo$ due to the presence of a DM core, here we use only those galaxies with $r_{\rm HI} > 2\re$. 

The stellar, dynamical, and halo masses for NGC1052-DF2 were obtained directly from \citet{vanDokkum18a}. The stellar and dynamical masses for NGC1052-DF4 were obtained from \citet{vanDokkum19}, and the limit on the halo mass was calculated by fitting a coreNFW profile \citep{Read16a} of average concentration to the 2$\sigma$ upper limit on the DM mass within $7\kpc$. Compared to a pure NFW profile, the inclusion of a DM core has a negligible effect on the halo mass obtained.

\section{Impact of parameter choices}
\label{sec:parameters}

\begin{figure*}
    \includegraphics[width=0.8\textwidth]{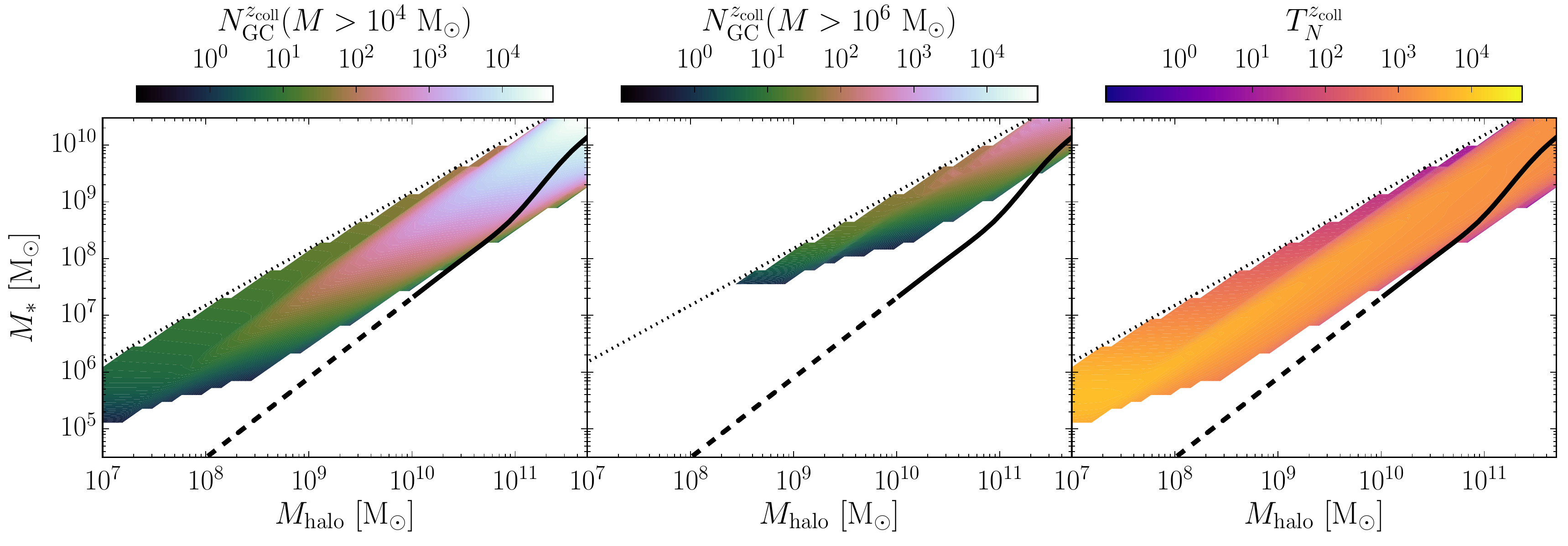}
    \includegraphics[width=0.8\textwidth]{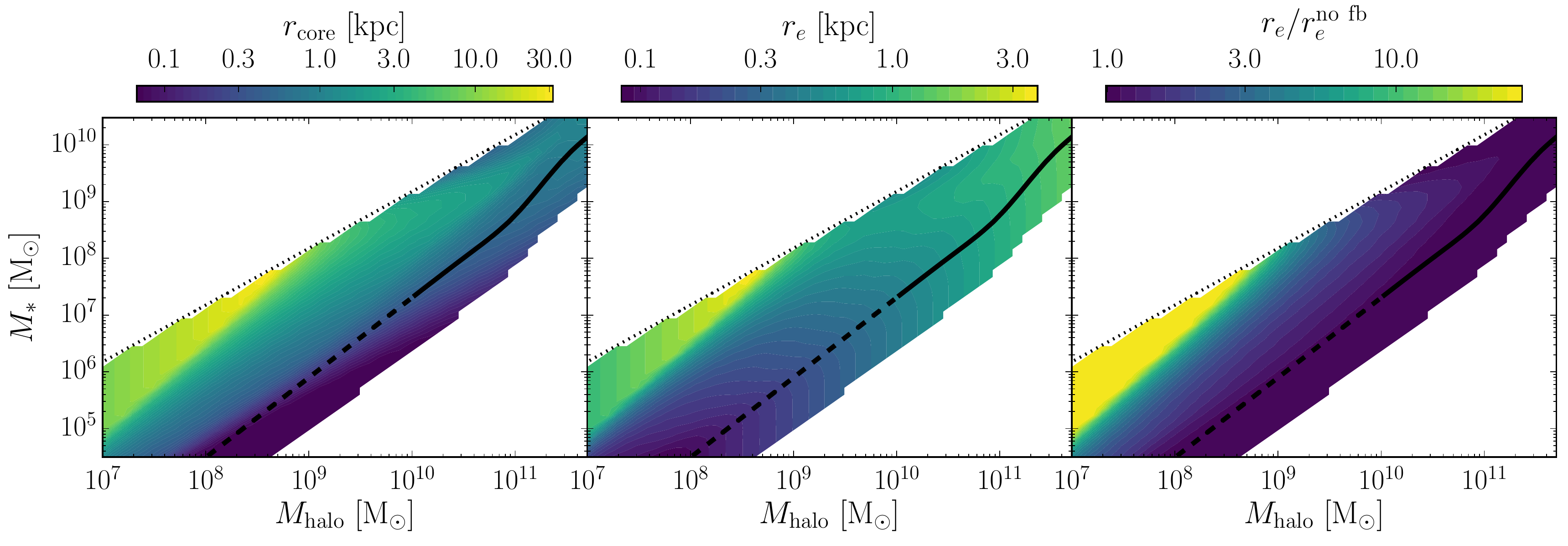}
    \caption{Predictions for GC populations and galaxy structure assuming a factor of 5 lower coupling of the wind energy to the DM, $\fDM=0.02$, compared to the fiducial value ($\fDM = 0.1$). Feedback-driven expansion of the DM halo and the galaxy is reduced due to the lower amount of energy available relative to the fiducial model (see Figs.~\ref{fig:global_GCs1}, \ref{fig:global_GCs2}, and \ref{fig:global_structure}). } 
\label{fig:global_fDM}
\end{figure*}

Here we examine the dependence of the predictions of the model on the choice of parameters. For this, we select the parameters that are least constrained by observations and that contribute the largest uncertainties. These are the fraction of wind energy that couples to the DM, $\fDM$, the fraction of stellar mass in place at $z=\zcoll$, $\fstar$, and the concentration - stellar mass relation, $c(\Mstar,\Mhalo)$. Figure~\ref{fig:global_fDM} shows the predictions for the GC populations and galaxy and halo structure for a model with the fiducial parameters and a 5 times lower value of the coupling fraction of wind energy to the DM, $\fDM=0.02$ (set to $\fDM=0.1$ in the fiducial model). As expected, the GC populations are unchanged, but the amount of expansion due to feedback is reduced by about a factor of $\sim 3$, while the maximum effect shifts to lower mass haloes with $\Mhalo\sim 10^{8.5}\Msun$.

\begin{figure*}
    \includegraphics[width=0.8\textwidth]{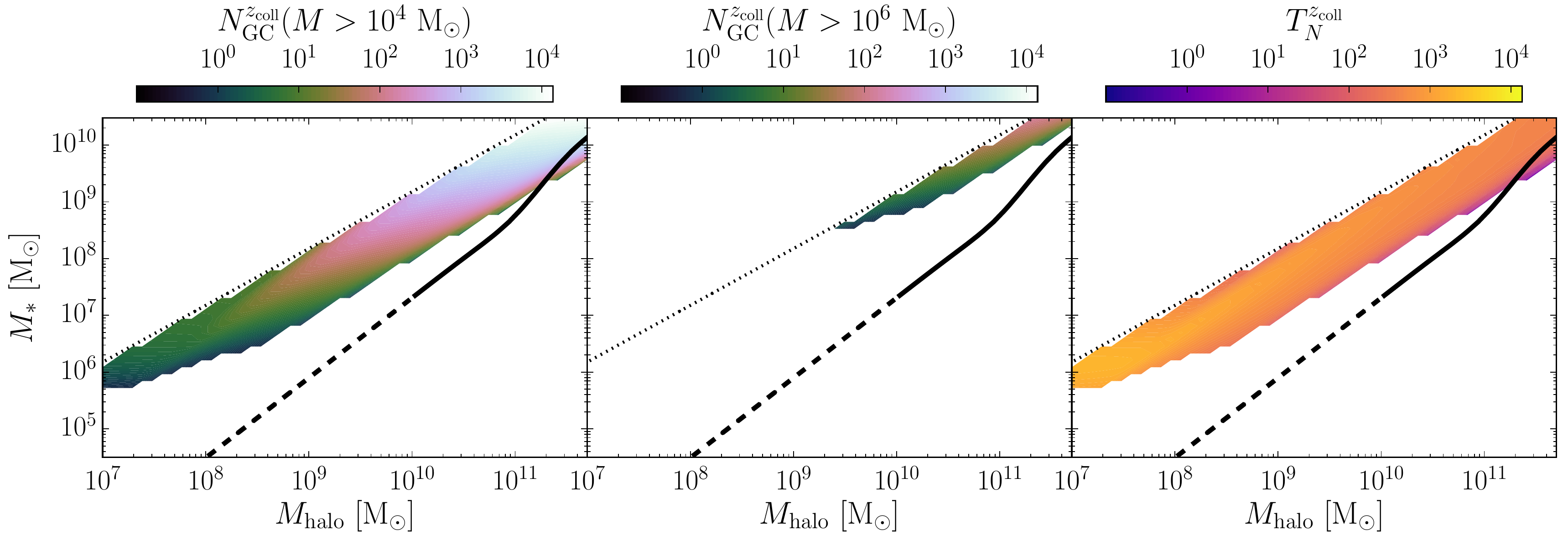}
    \includegraphics[width=0.8\textwidth]{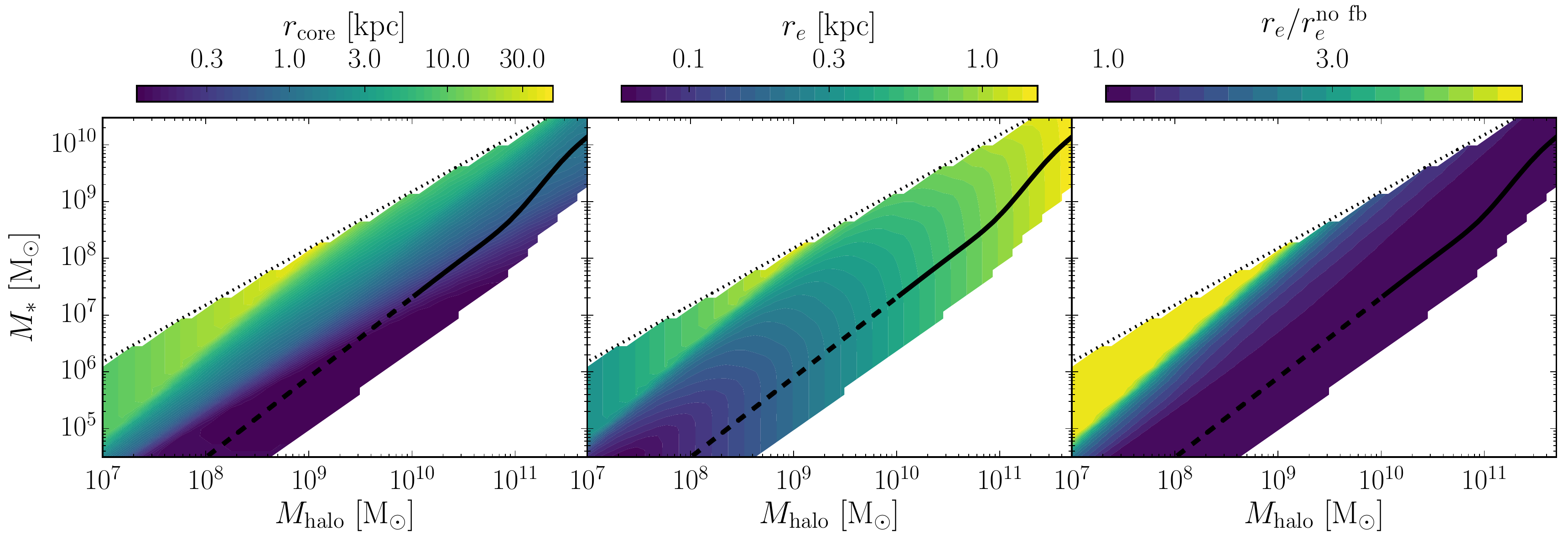}
    \caption{Predictions for GC populations and galaxy structure assuming a factor of 4 reduction in the fraction of stars formed at $z=\zcoll$, $\fstar=0.05$, relative to the fiducial value ($\fstar=0.2$). Due to the lower SFR at the collapse epoch, massive cluster populations are reduced in number, and the total SN energy is lower, leading to reduced feedback-driven expansion compared to the fiducial case (see Figs.~\ref{fig:global_GCs1}, \ref{fig:global_GCs2}, and \ref{fig:global_structure}). } 
\label{fig:global_fstar}
\end{figure*}

Figure~\ref{fig:global_fstar} shows the predictions for a model in which only 5 per cent of the present stellar mass of the galaxy formed by the collapse redshift $z=\zcoll$, $\fstar=0.05$, a factor of 4 lower than the fiducial value ($\fstar=0.2$). In this case, the number of massive clusters in galaxies near the mean SMHM relation is reduced by a factor of $\sim 4$, and the effect of expansion of the galaxy and halo shifts to lower halo masses $\Mhalo\la 10^{9}\Msun$ compared to the fiducial case.

\begin{figure*}
    \includegraphics[width=0.8\textwidth]{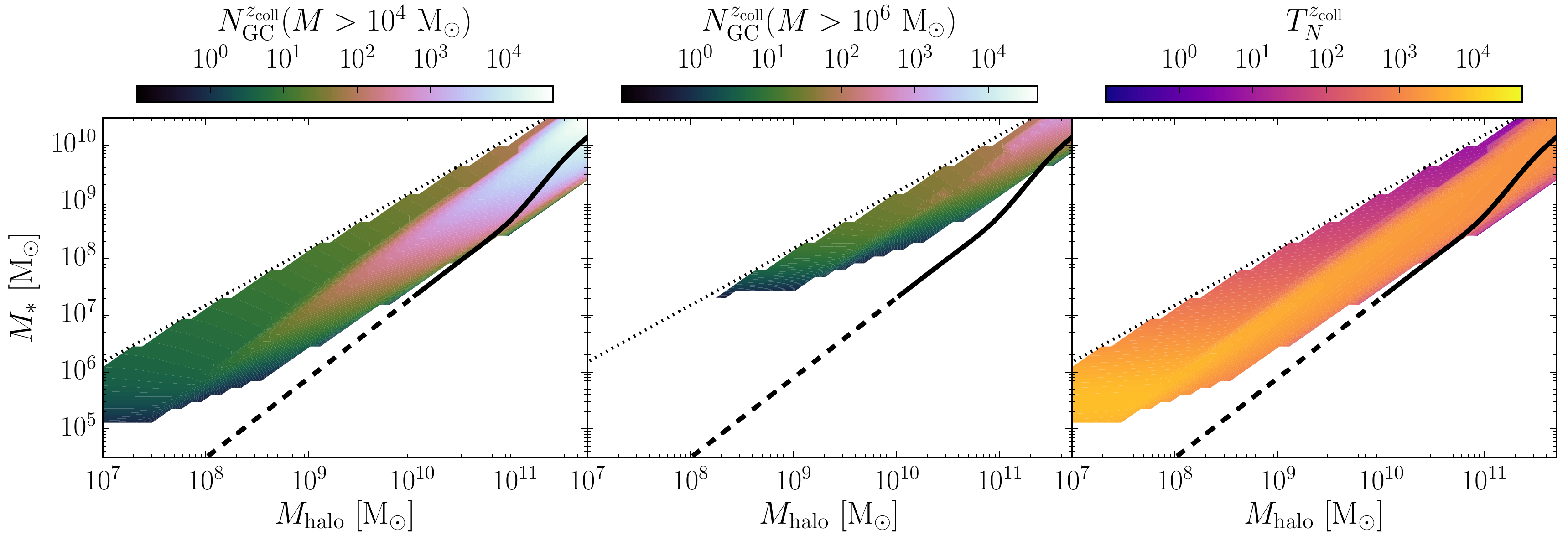}
    \includegraphics[width=0.8\textwidth]{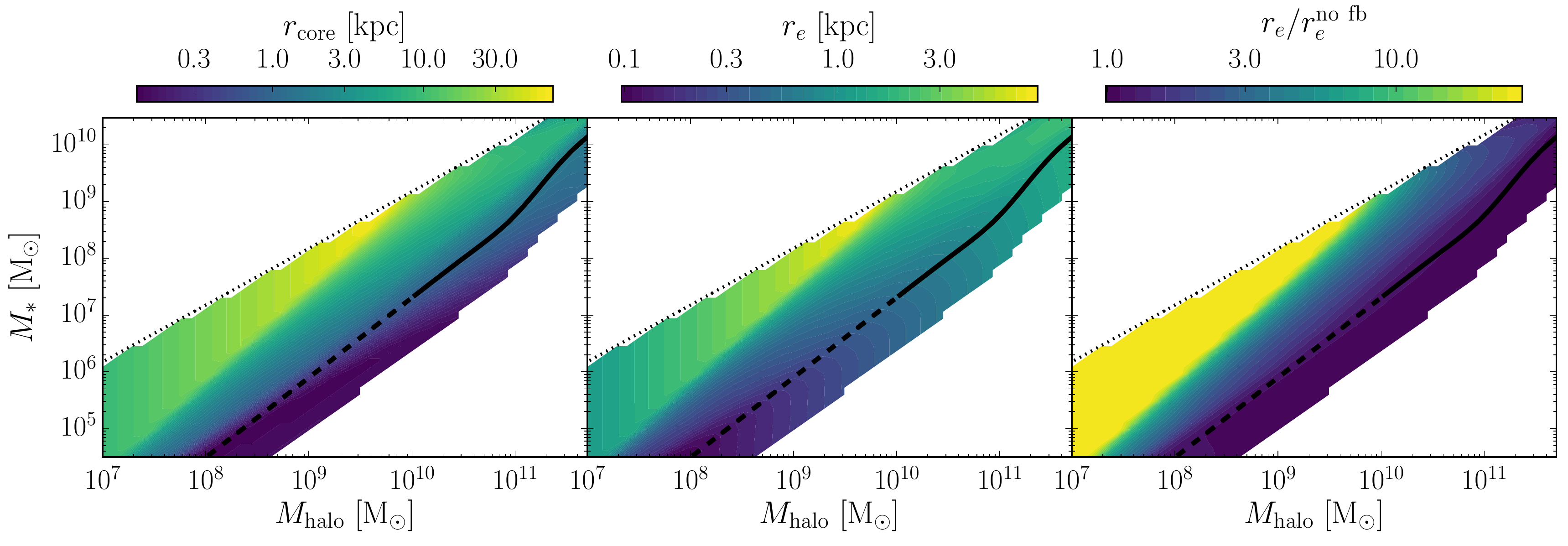}
    \caption{Global predictions for GC populations and galaxy structure assuming a steeper relation between halo concentration and offset from the mean SMHM relation (with $\fc=1.0$), relative to the fiducial model ($\fc=0.5$). Compared to the fiducial model (see Figs.~\ref{fig:global_GCs1}, \ref{fig:global_GCs2}, and \ref{fig:global_structure}), there is a slight reduction in the feedback-induced expansion of the galaxy and DM halo.  } 
\label{fig:global_c}
\end{figure*}

Figure~\ref{fig:global_c} shows the predictions for a model in which the relation between DM halo concentration and offset from the mean SMHM relation has a steeper slope by a factor of 2, $\fc=1.0$. This results in galaxies which scatter upwards having larger DM halo concentrations and earlier collapse times relative to the fiducial model (where $\fc=0.5$). The main consequence of the earlier collapse times is a slight reduction in the size of the GC populations of galaxies near the mean SMHM relation, and an earlier saturation for galaxies near the cosmic baryon fraction. The impact on galaxy and halo structure is minimal relative to the fiducial model.

\section{Influence of the Toomre $Q$ parameter}
\label{sec:toomre}

\begin{figure*}
    \includegraphics[width=0.8\textwidth]{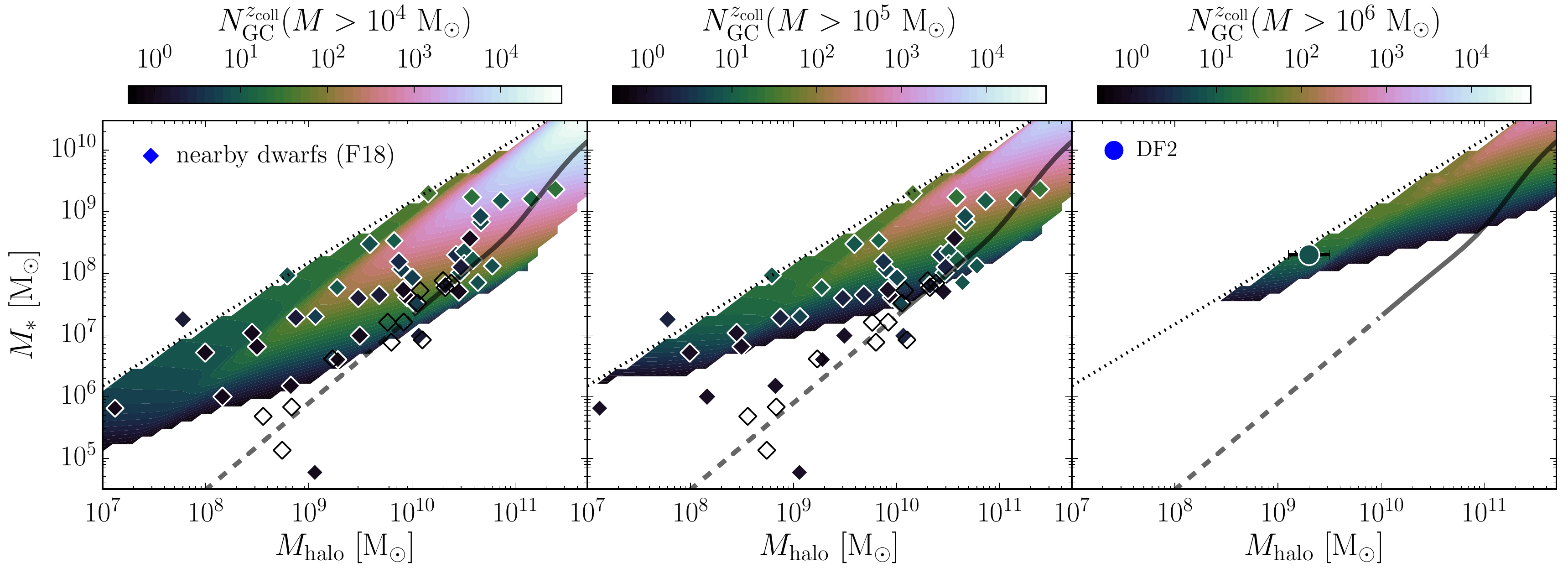}
    \includegraphics[width=0.8\textwidth]{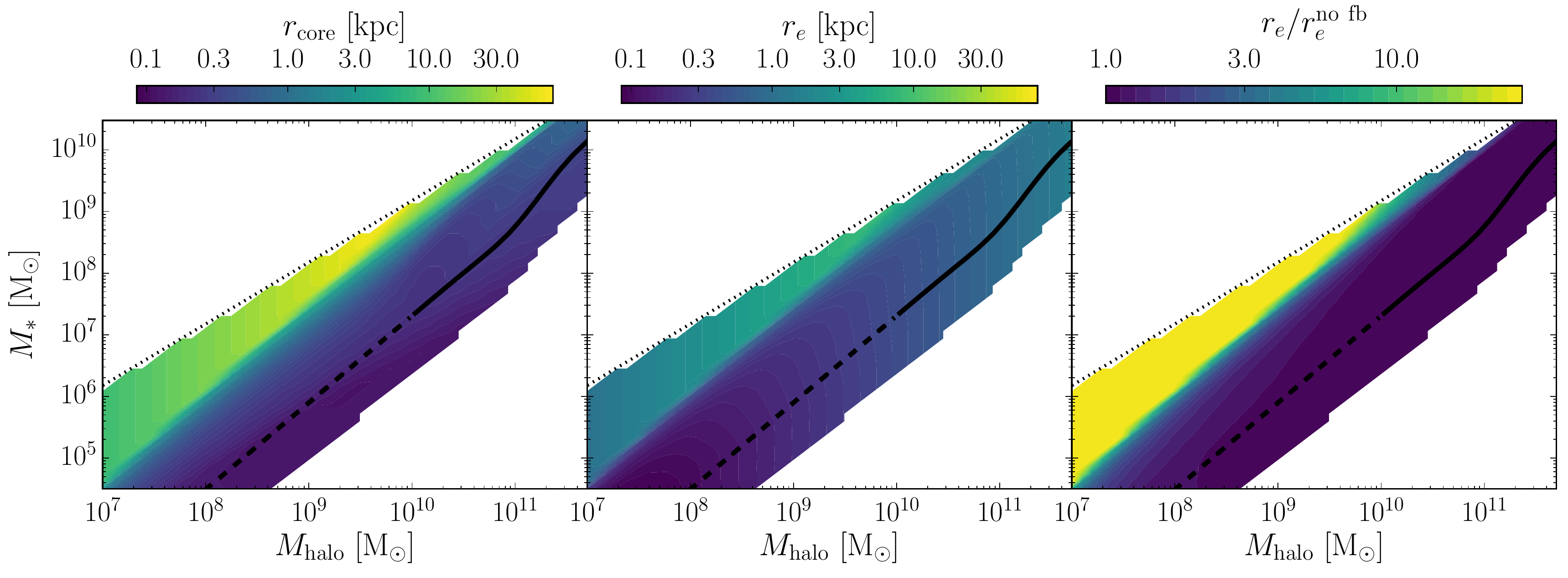}
    \caption{Predictions for GC populations and galaxy structure assuming a larger value of the Toomre parameter, $Q=2$ in the fiducial model. A larger $Q$ results in slightly larger GC populations for the most massive dwarf galaxies near the mean SMHM relation (see Fig.~\ref{fig:global_GCs1}). It also slightly reduces the region of UDG formation in the SMHM plane (see Fig.~\ref{fig:global_structure}).} 
\label{fig:global_Q}
\end{figure*}

Figure~\ref{fig:global_Q} shows the predictions for the fiducial model assuming a larger value of the Toomre parameter, $Q=2$, corresponding to the critical value for gravitational instability in a thin disc and $\fgas=0.5$. Larger values of $Q$ increase the largest unstable cloud scale and hence the maximum cluster mass, allowing for larger populations of massive GCs. Using this assumption, dwarf galaxies with very high specific frequencies like Fornax (located near the mean SMHM at $\Mhalo\approx 10^{10}\Msun$, and hosting 5 GCs) become typical in the model, with a prediction of $\NGC(M>10^{4}\Msun)\approx 14$. The model predictions for DF2/DF4 and the isolated gas-rich UDGs (see Section~\ref{sec:UDGs}) are able to fit the observational constraints as well as those with $Q=0.5$.

%%%%%%%%%%%%%%%%%%%%%%%%%%%%%%%%%%%%%%%%%%%%%%%%%%

% Don't change these lines
\bsp	% typesetting comment
\label{lastpage}
\end{document}